\documentclass{JHEP3}
\usepackage{amssymb}
\usepackage{amsfonts}
\usepackage{amsbsy}
\usepackage{amsmath}
\usepackage{amsthm}
\usepackage{graphicx}
\usepackage{epstopdf}
\usepackage[vcentermath]{youngtab}
\usepackage{multirow}

\newcommand{\ket}[1]{\vert #1 \rangle}
\renewcommand{\mod}[1]{\ (\mathrm{mod}\ #1)}

\def\M{\mathcal{M}}

\def\half{\frac{1}{2}}
\def\Z{\mathbb{Z}}
\def\F{\mathbb{F}}

\def\G{\mathcal{G}}

\def\cL{\mathcal{L}}
\def\R{\mathbb{R}}
\def\N{\mathcal{N}}
\def\P{\mathbb{P}}

\def\talpha{\tilde{\alpha}}
\def\nhv{H-V}
\def\n3a{t}

\def\tr{{\mathrm{tr}}}

\title{Global aspects of the space of 6D ${\cal N} = 1$ supergravities}

\author{Vijay Kumar$^{1, 2}$, David R.  Morrison$^3$ and Washington Taylor$^1$\\
$^1$Center for Theoretical Physics\\
Department of Physics\\
Massachusetts Institute of Technology\\
Cambridge, MA 02139, USA\\
\\
$^2$Kavli Institute for Theoretical 
Physics\\ University of California, Santa Barbara\\ Santa Barbara, CA 93106, USA\\
\\
$^3$Departments of Mathematics and  
Physics\\ University of California, Santa Barbara\\ Santa Barbara, CA 93106, USA\\
\\
{\tt vijayk} {\rm at} {\tt mit.edu},
{\tt drm} {\rm at} {\tt math.ucsb.edu},
{\tt wati} {\rm at} {\tt mit.edu}
}

\preprint{MIT-CTP-4132, NSF-KITP-10-096, UCSB-Math-2010-19}

\abstract{We perform
a
global analysis of the space of consistent 6D
quantum gravity theories with ${\cal N} = 1$ supersymmetry, including
models with multiple tensor multiplets.  We prove that for theories
with fewer than  $T = 9$ tensor multiplets, a finite number of distinct gauge
groups and matter content are possible.  We find infinite families of
field combinations satisfying anomaly cancellation and admitting
physical gauge kinetic terms for $T > 8$.  We find an integral lattice
associated with each apparently-consistent supergravity theory;
this lattice is determined by the form of the anomaly polynomial.  For
models which can be realized in F-theory, this anomaly lattice is related to
the intersection form on the base of the F-theory elliptic fibration.
The condition that a supergravity model have an F-theory realization
imposes constraints which can be expressed in terms of this
lattice.  The
analysis of models which satisfy
known low-energy consistency conditions and yet violate F-theory
constraints suggests possible novel constraints on low-energy
supergravity theories.}

\begin{document}

\parskip 0.1in

%\input{1}
%--------------------------------
\section{Introduction}

Six-dimensional ${\cal N} = 1$ supergravity 
is a useful domain for studying fundamental questions about the space
of string vacua.  On the one hand, the space of such theories is
strongly constrained by gravitational, gauge, and mixed anomalies
\cite{A-GW, gswest}.  On the other hand, 6D supergravity includes a rich
variety of models with different gauge groups and matter
representations.  Given the difficulty of attaining a systematic
classification of string vacua in four dimensions, and our limited
knowledge at this point regarding 
constraints on the space of low-energy 4D field theories which can be
consistently coupled to quantum gravity, 
it is desirable to find a context
in which we can develop tools and experience for addressing global
questions of this nature.  6D ${\cal
  N} = 1$ supergravity promises to provide a tractable framework in
which we can address questions such as the global extent of the space
of quantum-consistent low-energy theories, and begin to map how
different string vacuum constructions fill regions in this space of
low-energy theories.

In \cite{universality, finite, KMT} we began a systematic analysis of
the space of low-energy 6D ${\cal N} = 1$ supergravity theories.  
We showed in \cite{finite}
that for theories with one tensor field (those models which admit a
Lagrangian), there are a finite number of distinct nonabelian gauge
groups and matter representations possible
in models which
do not suffer from clear quantum inconsistencies from anomaly
violation or wrong-sign kinetic terms.  In \cite{KMT} we gave an
explicit map from the set of such models to topological data for
F-theory constructions.  For most apparently-consistent low-energy
models this map appears to give good string vacua through F-theory
constructions.  In some cases, however, the image of the map was found to
exhibit some pathology; in such cases the models do not 
correspond to any known F-theory vacuum.

In this paper, we extend this global analysis to theories with
multiple tensor fields.  The Lagrangian describing gauge groups and
matter fields for 6D ${\cal N} = 1$ supergravity theories with one
tensor field was originally developed in \cite{Nishino-Sezgin}.  Field
equations for the (non-Lagrangian) models with multiple tensor fields
were analyzed by Romans \cite{Romans}.  In a theory with $T$
antisymmetric 2-form tensor fields, the associated scalars
parameterize a coset space $SO(1, T)/SO(T)$.  We use the underlying
$SO(1, T)$ structure to give a fairly simple proof that for models
with $T < 9$, there are a finite number of possible nonabelian gauge
groups and matter representations consistent with anomaly cancellation
and physical gauge kinetic terms.  We find that for any $T$, each
apparently-consistent low-energy supergravity theory can be associated
with an integral lattice $\Lambda$.  We use the structure of the
lattice $\Lambda$ to connect with F-theory, and identify constraints
on low-energy theories associated with the existence of an F-theory
vacuum construction.  Thus, the results of this paper generalize and
subsume many of the results of \cite{finite, KMT}; the general
structure developed for models with arbitrary $T$ clarifies in some
ways the more specific arguments previously given at $T = 1$.

In Section \ref{sec:anomalies} we review the structure of anomaly
cancellation in general 6D ${\cal N} = 1$ supergravity theories, and
demonstrate that each anomaly-free theory can be associated with an
integral lattice.  In Section \ref{sec:finite} we prove that the
number of gauge groups and matter representations for consistent
theories is finite for $T < 9$.  We give explicit examples of infinite
families where the theorem breaks down at $T \geq 9$ in Section \ref{sec:examples-infinite}.
In Section \ref{sec:F-theory} we use the
integral lattice for each theory to construct topological data which
would be associated with any corresponding F-theory construction, and
describe constraints on the class of models which can be realized
through F-theory.  Section \ref{sec:examples} contains some
examples  of F-theory embeddings of supergravity models, illustrating
how geometrical constraints from F-theory rule out the 
possibility of
F-theory realizations of some apparently-consistent
low-energy models.
Section \ref{sec:global} contains some discussion of global aspects of
the 6D supergravity landscape, and Section \ref{sec:conclusions}
contains a summary of the conclusions.

Note that
the analysis in Sections \ref{sec:anomalies}, \ref{sec:finite},  and
\ref{sec:examples-infinite} depends
only upon the structure of low-energy supergravity and is independent
of any structure associated with string theory.

\section{Anomalies and lattices}
\label{sec:anomalies}

We consider 6D supergravity theories with semi-simple gauge group $\G
= \prod_i G_i$.  The analysis of  abelian factors
will appear elsewhere \cite{Park-Taylor}; such abelian factors have
little effect on the structure of the nonabelian part of the theory.
We include matter hypermultiplets which transform in a general
representation of $\G$, and $T$ tensor multiplets.  For theories with
multiple tensor multiplets, there is a generalized Green-Schwarz
mechanism
(described by
Sagnotti \cite{Sagnotti}), which allows for a larger class of gauge
anomalies to be cancelled.  In this section we review this mechanism
using the notation of \cite{Sadov}, and show that the anomaly
cancellation conditions imply the existence of an integral lattice
associated with any consistent 6D  ${\cal N} = 1$ theory.

\subsection{Anomaly cancellation}
Anomalies can be cancelled by the Green-Schwarz-Sagnotti mechanism if
the anomaly polynomial $I_8$ can be written in the form
\begin{equation}
I_8(R,F) = \half \Omega_{\alpha\beta} X^\alpha_4 X^\beta_4
\label{eq:factorized-anomaly}
\end{equation}
Here $X^\alpha_4$ is a 4-form constructed from the curvatures of the Yang-Mills and spin connections
\begin{equation}
X^\alpha_4 = \half a^\alpha \tr R^2 +  \sum_i b_i^\alpha \ 
\left(\frac{2}{\lambda_i} \tr F_i^2 \right)
\end{equation}
where $a^\alpha, \ b_i^\alpha$ are vectors in the space $\R^{1,T}$ and
$\Omega_{\alpha\beta}$ is a natural metric (symmetric bilinear form)
on this space.
The
$\tr$ here refers to the trace in an appropriate ``fundamental''
representation of the group $G_i$, and $\lambda_i$ is a normalization
factor.  These normalization factors
are fixed by demanding that the smallest topological charge of an embedded
$SU(2)$ instanton is 1,
as explained in \cite{Guth-etal}.  These factors are listed in Table
\ref{table:norm} for all the simple groups.  $I_8(R,F)$ is completely
specified by the multiplets in the low-energy theory and can be
computed using the formulae in \cite{gsw2, Erler}.
\begin{table}[b]
\centering
\begin{tabular}{|c|c|c|c|c|c|c|c|c|c|}
\hline
        & $A_n$ & $B_n$ & $C_n$ & $D_n$  & $E_6$ & $E_7$ & $E_8$ & $F_4$ & $G_2$ \\ 
        \hline
$\lambda$ & $1$   & $2$ & $1$ &$2$ & $6$ & $12$ & $60$ & $6$ & $2$ \\
\hline
\end{tabular}
\caption{Normalization factors for the simple groups}
\label{table:norm}
\end{table}

When $T \neq 1$, one cannot write down a Lorentz covariant Lagrangian.
One can construct a partition function, however, along the lines of
\cite{Witten-M5-brane, Belov-Moore} by coupling the extra (anti-self-dual) tensor
fields to an auxillary 3-form gauge potential.  
The anomaly can be cancelled by a local counterterm of the form
\begin{equation}
\delta \cL_{GSS} = -\Omega_{\alpha \beta} B^\alpha \wedge X^\beta_4
\end{equation}
The 2-form
field $B^\alpha$ has an anomalous gauge transformation, and the above term makes the tree-level Lagrangian gauge-variant in exactly the right way to cancel the one-loop anomaly.  The gauge invariant 3-form field strength is defined as
\begin{align}
H^\alpha  = dB^\alpha + \half a^\alpha \omega_{3L} + 2 \sum_i \frac{1}{\lambda_i} b_i^\alpha  \omega^i_{3Y}
\end{align}
where $\omega_{3L}$ and $\omega_{3Y}^i$ are Chern-Simons 3-forms of the spin connection and gauge field respectively.

The only $(1,0)$ supersymmetry multiplets that contain scalars are the
hypermultiplets (4 real scalars) and the tensor multiplet (1 real
scalar).  The hypermultiplet moduli space is a quaternionic K\"ahler
manifold, analogous to the 4D $\N=2$ case.  The tensor multiplet
scalar
moduli space locally takes the form of the coset space $SO(1,T)/SO(T)$
\cite{Romans}, and can be parameterized by a vector $j^\alpha$ in the
space $\R^{1,T}$ of norm $\Omega_{\alpha\beta}j^ \alpha j^\beta=+1$.
From the viewpoint of the low-energy
theory, the above space is just one possible solution to the
requirements imposed by SUSY, although no others are known.  In Section
\ref{sec:geometry} we will see that the space $SO(1,T)/SO(T)$ arises
naturally as a coset space containing
the Teichmuller space of K\"ahler metrics in F-theory
compactifications as an open set.

The anomaly polynomial does not specify the vectors
$a^\alpha, \ b_i^\beta$, but only constrains the $SO(1,T)$ invariant
quantities
\begin{equation}
\Omega_{\alpha \beta} a^\alpha a^\beta, \quad \Omega_{\alpha\beta} a^\alpha b_i^\beta, \quad \Omega_{\alpha\beta}b_i^\alpha b_j^\beta
\end{equation}
We will use the notation $x\cdot y$ to denote the $SO(1,T)$ invariant
product $\Omega_{\alpha\beta} x^\alpha y^\beta $.  
Vanishing of the $\tr R^4$ anomaly implies that
\begin{equation}
H-V = 
273- 29T
\label{eq:bound}
\end{equation} 
where $H, V, T$ denote the number of hyper, vector, and tensor
multiplets respectively.
The $\tr F^4$ contribution to the total anomaly must also cancel for the
anomaly to factorize in the form (\ref{eq:factorized-anomaly}).  This gives the condition
\begin{equation}
 B^{i}_{\rm adj}  =  \sum_R x^i_{R} B^i_{R} \label{eq:f4-condition}
\end{equation}
where the coefficients $A_R,B_R,C_R$ are defined as
\begin{align}
\tr_R F^2 & = A_R  \tr F^2 \\
\tr_R F^4 & = B_R \tr F^4+C_R (\tr F^2)^2
\end{align}
and 
$x_R^i$
denotes the number of hypermultiplets transforming in
representation $R$ under gauge group factor $G_i$.
We will similarly denote by $x^{ij}_{RS}$ the number of
hypermultiplets transforming in representations $R, S$ under the
factors $G_i, G_j$.

The remaining anomaly factorization conditions relate inner
products between the vectors $a, b_i$ to group theory coefficients and
the representations of matter fields
\begin{align}
a \cdot a & =  9 - T \\ 
a \cdot b_i & =  \frac{1}{6} \lambda_i  \left( A^i_{\rm adj} - \sum_R
x^i_R A^i_R\right)  \label{eq:ab-condition}\\
b_i\cdot b_i & = -\frac{1}{3} \lambda_i^2 \left( C^i_{\rm adj} - \sum_R x_R^i C^i_R  \right) \\
b_i \cdot b_j & =  \lambda_i \lambda_j \sum_{RS} x_{RS}^{ij} A_R^i
A_S^j.
\label{eq:bij-condition}
\end{align}
We demonstrate in the following section that these inner products are
all integers, so that $a, b_i$ can be used to define an integral lattice.

In the case when $T=1$, which was studied in \cite{gswest,
  6D-anomalies, finite}, the Green-Schwarz mechanism requires that the
anomaly polynomial factorize
as a simple product of polynomials.
To relate this familiar case to the
general formalism, we can choose the bilinear form on $SO(1,1)$ as
\begin{equation}
\Omega_{\alpha\beta} = \begin{pmatrix}
0 & 1 \\
1 & 0
\end{pmatrix} 
\label{eq:canonical-omega}
\end{equation}
For $T = 1$ we have $a^2 = 8$.
The form (\ref{eq:canonical-omega}) and the anomaly polynomial (\ref{eq:factorized-anomaly}) are invariant under the rescaling 
\begin{equation}
(X_4^1, X_4^2) \rightarrow (\mu \ X_4^1, \mu^{-1}\  X_4^2)
\end{equation}
We can use this scale degree of freedom to set $a^\alpha\equiv (a^1,a^2)=(-2,-2)$.  
We can then identify $b_i$ in this basis with parameters $\alpha_i,
\tilde{\alpha}_i$ through
\begin{equation}
b_i = \frac{\lambda_i}{2}(\alpha_i, \tilde{\alpha}_i) \,.
\label{eq:ba}
\end{equation}
Once this basis is chosen for $\Omega_{\alpha\beta}, a,$ and $b_i$, the anomaly
polynomial takes the familiar, friendly form used in \cite{KMT} and
most literature on $T = 1$ models
\begin{equation}
I_8 = X^1 X^2 =
(\tr R^2 - \sum_i \alpha_i \tr F_i^2)(\tr R^2 - \sum_i \talpha_i \tr F_i^2)
\end{equation}
The only
symmetry unfixed is a $Z_2$ symmetry of the bilinear form that
exchanges $X^1 \leftrightarrow X^2$.

\subsection{Proof of integrality}

The inner products of the vectors $a^\alpha,\ b_i^\alpha$ are related to 
group invariants ($A_R,\ C_R$) and the number of charged hypermultiplets in 
various representations. 
In this section we show that these inner 
products are quantized in $\Z$ when the normalization factors $\lambda_i$ 
are chosen as in Table \ref{table:norm}.  
(\ref{eq:ab-condition}-\ref{eq:bij-condition}) imply that the inner products $a\cdot b_i, \ b_i \cdot b_i, \ b_i \cdot b_j$ are all quantized in integers if
\begin{eqnarray}
  \lambda_i  \frac{\sum_R x^i_R A^i_R - A^i_{\rm adj}}{6} \in \Z    \nonumber \\
  \lambda_i^2 \frac{\sum_R x^i_R C^i_R - C^i_{\rm adj}}{3} \in \Z   \label{eq:int-conditions}\\
  \lambda_iA_R^i \in \Z  \nonumber 
\end{eqnarray}
We will prove the above statements for each of the simple groups, case by 
case.  For the $SU(N)$ (and $Sp(N)$) series, this is easily proved using 
properties of Young diagrams, which are in 
one-to-one
correspondence with 
the irreducible representations of $SU(N)$ (and $Sp(N)$).  For an arbitrary 
group $G$
not of the form $SU(N)$ or $Sp(N)$,
we find a sequence of maximal subgroups that  terminates in 
$SU(N)$ or $Sp(N)$, e.g.  $E_8 \supset SU(9), \ SO(8) \supset SO(7) 
\supset SO(6) \cong SU(4)$.  Then, by computing the branching of 
representations of $G$ we can show integrality for $G$.

We start with $SU(N)$, $N \geq 4$; the coefficients $A_R, B_R, C_R$ can 
be easily 
calculated using two diagonal generators $T_{12}, \ T_{34}$ 
which, in the fundamental representation, take the form
\begin{align}
(T_{12})_{ab} &= \delta_{a1}\delta_{b1}-\delta_{a2}\delta_{b2} \\
(T_{34})_{ab} & = \delta_{a3}\delta_{b3} - \delta_{a4}\delta_{b4}
\end{align}
as in the 
appendix of \cite{finite}.
The group theory factors $A_R, \ B_R, \ C_R$ can be computed in terms
of traces of these generators.
\begin{align}
A_R & = \frac{1}{2} \tr_R T_{12}^2  \label{eq:A-def}\\
B_R+2 C_R & = \half \tr_R T_{12}^4 \label{eq:B-def}\\
C_R & = \frac{3}{4} \tr_R T_{12}^2T_{34}^2 \label{eq:C-def}
\end{align}
In the traces above, we sum over all states in the 
representation $R$, which can be represented in terms of the
associated Young diagram $D_R$.  
In
(\ref{eq:C-def}), let $\ket{s}$ denote a state in the representation normalized to $\langle s | s \rangle=1$.
A basis of such states corresponds to the set of Young tableaux,  given by the Young
diagram $D_R$
labelled using integers from $1, 2, \cdots, N$.  Let $\pi_{ij}$ denote the operation
that switches the labels $i \leftrightarrow j$ in a Young tableau.  The
states $\{\ket{s}, \pi_{12}\ket{s}, \pi_{34} \ket{s}, \pi_{12}\circ
\pi_{34}\ket{s}\}$ all give equal contributions to the trace.  
(Note that these states need not be distinct, but when they are not
all distinct, the contribution vanishes, since either the number of
appearances of 1 and 2 are equal or the number of appearances of 3 and
4 are equal.)
This
makes $\tr_R T_{12}^2T_{34}^2$ a multiple of 4, and therefore $C_R$ is
an integer divisible by 3 for every representation of $SU(N), \ N \geq
4$.  This shows that for $SU(N), N \geq 4$,
\begin{equation}
\frac{1}{3} \left( \sum_R x_R C_R - C_{\rm adj} \right) \in \Z
\label{eq:C-div}
\end{equation}
Anomaly cancellation requires the vanishing of the $\tr F^4$ term,
which sets $\sum_R x_R B_R - B_{\rm adj}=0$.  If we define $E_R:=\half
\tr_R T_{12}^4$, then (\ref{eq:B-def}), (\ref{eq:C-div}) 
and the $F^4$ condition
together imply that $\sum_R x_R
E_R-E_{\rm adj}$ is divisible by 6.  From the
same kind of
argument as above we know that
in a representation $R$, the states $\ket{s}, \ \pi_{12} \ket{s}$
would 
together
contribute $n(s)^2$ to $A_R$ and $n(s)^4$ to $E_R$, with $n(s) :=
\langle s | T_{12} \ket{s} \in \Z$.  Since $n(s)^2 \equiv n(s)^4
\mod{6}$, we have $E_R \equiv A_R \mod{6}$.  Therefore,
\begin{equation}
\frac{1}{6}\left(\sum_R x_R A_R - A_{\rm adj}\right) \in \Z \,.
\end{equation}
Since the normalization factor for $SU(N)$ in Table \ref{table:norm} is 1, this shows that for an $SU(N \geq 4)$ factor in the gauge group all the conditions in (\ref{eq:int-conditions}) are satisfied.

For the other subgroups of $GL(N)$, we can again use the Young diagram
approach (see \cite{Hamermesh} for details on Young diagrams for the other classical groups).  
The irreducible
representations of $Sp(N)$ are in one-to-one correspondence with Young
diagrams with all possible contractions with the $\epsilon$ symbol
subtracted.  If we choose appropriate generators from the Cartan
subalgebra of $Sp(N)$ as in \cite{finite}, whose squares are exactly
equal to the squares of the $SU(N)$ generators we chose above, the
above proof for $SU(N)$ carries through unchanged.  In the case of
$SO(N)$, the set of Young diagrams only gives the representations
with integer weight; these exclude the spinor
representations.  With the right choice of generators, the above proof
shows that for the integer weight representations, we have integral
inner products.  If we include spinor representations, the inner
products are actually quantized in $\half \Z$.  
Including the
factor of 2 in the normalization of $SO(N)$ in Table \ref{table:norm},
however,
we find that the inner products are again integral.  
An alternate proof of
integrality for the $SO(N)$ representations involves using a sequence
of maximal subgroups.  We will discuss this method in general below,
and then apply it to the $SO(N)$ case and also to the exceptional
groups.

Consider a general simple group $G$.  We wish to show that whenever
$\sum_R x_R B_R - B_{\rm adj}=0$, the conditions (\ref{eq:int-conditions})
are satisfied.  Let $H$ be a simple, maximal (proper) subgroup of the
simple Lie group $G$.  For a given representation $R$ of $G$, we can
compute the decomposition of $R$ into irreducible representations
$S_i$ of the maximal subgroup $H \subset G$: $R = \oplus_i n(R)_i S_i$
where $n(R)_i$ denotes the multiplicity of representation $S_i$.
\begin{align}
\tr_R F_G^2 & = \sum_i n(R)_i \tr_{S_i} F_H^2 = \left(\sum_i n(R)_i A_{S_i}(H)\right) \tr F_H^2\\
\tr_R F_G^4 & = \sum_i n(R)_i \tr_{S_i} F_H^4 = \left(\sum_i n(R)_i B_{S_i}(H) \right) \tr F_H^4 + \left(\sum_i n(R)_i C_{S_i}(H) \right) (\tr F_H^2)^2
\end{align}
Here $\tr$ denotes the trace in the fundamental representation of $H$.
This gives us a way of computing $A_R(G)$ for an arbitrary group $G$ using its maximal subgroup \cite{Erler}.  
\begin{align}
A_R(G) = \frac{\tr_R F_G^2}{\tr_f F_G^2} = \frac{\sum_i n(R)_i
  A_{S_i}(H)}{\sum_i n(f)_i A_{S_i}(H)} \,.
\end{align}
Here, for clarity, $\tr_f$ explicitly denotes the trace in the (suitably normalized) fundamental representation $f$ of $G$ with $f = \oplus_i n(f)_i S_i$ under $H\subset G$.

Anomaly cancellation for $G$ implies that $\sum_R x_R B_R(G) - B_{\rm adj}(G) = 0$.  Then, 
\begin{align}
\sum_R x_R \tr_R F^4_G - \tr_{\rm adj} F^4_G & = \left(\sum_R x_R C_R(G) - C_{\rm adj}(G)
\right) (\tr_f F_G^2)^2 \nonumber \\
\Rightarrow \sum_R x_R C_R(G) - C_{\rm adj}(G) & = \frac{\sum_R x_R \sum_i n(R)_i C_{S_i}(H) - \sum_i n(adj)_i C_{S_i}(H)}{(\sum_i n(f)_i A_{S_i}(H))^2} \,.
\end{align}
If we have integrality for the group $H$, i.e.  for every $\{y_S \in \Z\}$ satisfying 
\begin{equation}
\sum_S y_S B_S(H)=0 \Rightarrow  \left\{  
\begin{array}{c}
\lambda_H \sum_S y_S A_S(H) \in 6 \Z \\
\lambda_H^2\sum_S y_S C_S(H)\in 3 \Z
\end{array} \right.
\end{equation}
then, using $y_{S_i} = \sum_R x_R\, n(R)_i -n ({\rm adj})_i$ we have
\begin{align}
\lambda_H\left(\sum_i n(f)_i A_{S_i}(H)\right) \frac{\sum_R x_R A_R(G) - A_{\rm adj}(G)}{6} \in \Z \\
\lambda_H^2 \left(\sum_i n(f)_i A_{S_i}(H)\right)^2 \frac{\sum_R x_R C_R(G) - C_{\rm adj}(G)}{3} \in \Z
\end{align}
The conditions (\ref{eq:int-conditions}) are then all satisfied for a group $G$, if we prove that 
\begin{equation}
\lambda_G = \lambda_H \sum_i n(f)_i A_{S_i}(H) \label{eq:int-GH}
\end{equation}
for a maximal subgroup $H$.

We first consider the particular case of $G=E_8$ and the maximal subgroup $SU(9)\subset E_8$ \cite{Erler}.  The coefficients $A_R,B_R,C_R$ for $E_8$ are defined using the adjoint representation as the fundamental.  
\begin{align}
E_8 & \supset SU(9)  \nonumber \\
248 & = 84 \ \oplus \ \overline{84}\ \oplus \  {\rm adj}  \label{eq:E8-branch}
\end{align}
Using $A_{84}(SU(9)) = 21, \ A_{\rm adj}(SU(9))=18$, we have integrality
for $E_8$, because $\lambda_{E_8} = 60$ and $\lambda_{SU(9)}=1$, and
therefore relation (\ref{eq:int-GH}) is satisfied
\begin{equation}
\lambda_{E_8}  = \lambda_{SU(9)} ( 2A_{84} + A_{\rm adj} )  = 60\,.
% \label{eq:}
\end{equation}
In fact, the branching rule in (\ref{eq:E8-branch}) makes clear the origin of the normalization factors. Since
\begin{equation}
\tr_{\rm adj} F_{E_8}^2 = 2 \ \tr_{84} F_{SU(9)}^2 + \tr_{\rm adj} F_{SU(9)}^2 = 60\ \tr_9 F_{SU(9)}^2, 
\end{equation}
including a normalization factor of  1/60 for the $E_8$ group trace, relative to the $SU(9)$ trace,  ensures that the minimum instanton number of any configuration of $E_8$ gauge fields is 1.

For the $SO(N)$ series, we can show that (\ref{eq:int-GH}) is satisfied by induction.  For $N = 6$, $SO(6) \cong SU(4)$ we have
\begin{equation}
\tr_6 F_{SO(6)}^2 = \tr_6 F_{SU(4)}^2 = 2 \,\tr F_{SU(4)}^2
\end{equation} 
Equation (\ref{eq:int-GH}) is again satisfied since $\lambda_{SO(6)}=2, \lambda_{SU(4)}=1$.  For the inductive step, since $SO(N-1)$ is a maximal subgroup of $SO(N)$,
\begin{equation}
\tr_N F_{SO(N)}^2 = \tr_{N-1} F_{SO(N-1)}^2
\end{equation}
Thus, integrality for $SO(N-1)$ implies integrality for $SO(N)$, and the inductive step is proved.

Similarly, we can prove integrality for the groups $E_6, \ E_7, \ F_4$
using the maximal subgroups $Sp(4), \ SU(8), \ SO(9)$ respectively.
In each case we find that relation (\ref{eq:int-GH}) is
satisfied.
We have thus shown that the inner products $a\cdot b_i, \ b_i\cdot
b_i, \ b_i \cdot b_j$ are all integral
if all simple factors in the gauge group are drawn from the list
\begin{equation}
\{SU(N \geq 4), \ SO(N \geq 7), \ Sp(N \geq 2),  \ E_{6,7,8}, F_4 \}
\end{equation}
and suitable scaling factors are applied to the anomaly coefficients.
The groups $SU(2), SU(3)$ and $G_2$ are conspicuously absent from this
list.  The normalization factors for these groups are 1, 1 and 2
respectively, but in these cases, the condition that local anomalies
are absent does not constrain the inner products to be integral.
There is a more subtle anomaly, however, first discussed in
\cite{witten-global-anomaly}, where the partition function is
invariant under local gauge transformations (gauge current is
conserved quantum mechanically), but not invariant under ``large''
gauge transformations.  The analysis of such ``global anomalies'' in
six dimensions was carried out in \cite{Bershadsky-Vafa}, and more
thoroughly in \cite{Suzuki-Tachikawa}. Using their results, we find
that the inner products in question are non-integral for $SU(2),
SU(3)$ and $G_2$ precisely when the low-energy theory is plagued by a
global anomaly, which renders these theories inconsistent.

Therefore, imposing that the low-energy theory is free of local and global anomalies, we have shown that the anomaly coefficients define an integral lattice $\Lambda$.  This lattice $\Lambda$ will play a crucial role in defining possible embeddings into F-theory.

\subsection{Integral lattices and dyonic strings}
\label{sec:lattice}

Since the inner products $a \cdot a, a \cdot b_i, b_i \cdot b_j$
compatible with the anomaly cancellation equations are all integral,
we can use this inner product structure to form an integral lattice
\begin{equation}
\Lambda = \left(
\begin{array}{cccc}
a^2 & -a \cdot b_1 & -a \cdot b_2 & \cdots\\
-a \cdot b_1 & b_1^2 & b_1 \cdot b_2 & \cdots\\
-a  \cdot b_2 & b_1 \cdot b_2 & b_2^2 & \cdots\\
\vdots & \vdots & \vdots & \ddots
\end{array} \right)
\label{eq:lattice}
\end{equation}
Note that this lattice may be degenerate; in some cases there can be
linear relations between the vectors $a, b_i$.
We choose to define the lattice in terms of $-a$ rather than $a$ since
generally  $-a$ is a positive vector in the sense that $-a \cdot j >
0$ for those models with F-theory descriptions, as we discuss in
Section \ref{sec:F-theory}.

For models which have a consistent quantum UV completion, there is a
natural interpretation of the lattice $\Lambda$ in terms of the charge 
lattice of BPS states.  The BPS states of the 6D $\N =(1,0)$ SUSY 
algebra are extended string-like states, known as dyonic strings,  
with arbitrary charges under the $(1, T)$ multiplet of two-form fields 
$B^\alpha$ \cite{Duff-dyons}.  
For theories with nonabelian gauge fields, there are BPS states known 
as gauge dyonic strings, where the gauge field has an instanton 
profile in the directions transverse to the string \cite{Duff-gauge}.
For every nonabelian factor $G_i$ in the gauge group, there is a 
corresponding gauge dyonic string with conserved 2-form charge given 
by the vector $b_i$.  (The dyonic string is obtained from the gauge 
dyonic string by taking the instanton size $\rightarrow  0 $ limit.).  
In a consistent quantum theory, just as the
product of electric and magnetic monopole charges is quantized in
standard 4D electromagnetic theory due to the single-valued nature of
the electron wave function, the inner product $b \cdot b'$ of dyonic
string charges is quantized in the 6D theory \cite{Deser-quant}.  Thus, we expect that in
a consistent quantum theory, if there are quantum excitations
associated with the solitonic dyonic strings, these states must live
in an integral lattice $\tilde{\Lambda}$ of signature $(1, T)$ of
which $\Lambda$ is a sublattice.  It is interesting and perhaps
suggestive that the integrality of the lattice $\Lambda$ follows
directly from the anomaly cancellation conditions, with no further
assumptions about the quantum consistency of the theory or the
existence of quantum string states.

\section{Finite bound  for fixed $T < 9$}
\label{sec:finite}

In this section we prove that for fixed $T < 9$ there are a finite number
of distinct possible combinations of nonabelian gauge group and matter
representations.
This analysis is purely based on aspects of the low-energy
supergravity theory, and is independent of string theory or any other
specific UV completion.

The finite range of possible gauge groups and matter representations
for $T = 1$ was proven in \cite{finite}.  We give a similar proof here
for any fixed $T$ between 0 and 8, using the $SO(1, T)$ invariant
inner product structure on the vectors $a, b_i, j$.  As in
\cite{finite}, we ignore abelian factors; such factors do not affect
the anomaly cancellation conditions
(\ref{eq:f4-condition}-\ref{eq:bij-condition}) on the nonabelian gauge
group factors.  The constraint on infinite families breaks down at $T
= 9$ due to the change of sign of $a^2 = 9-T$.  When $a$ has positive
norm, it places stronger constraints on the range of allowed models.
We give explicit examples of  infinite families of anomaly-free models
with acceptable gauge kinetic terms at $T = 9$ and greater in Section
\ref{sec:examples-infinite}.

The proof for $T < 9$ proceeds by contradiction.  We assume that there
is an infinite family of models $\{\M_{(\gamma)}\} =  \{{\cal M}_{(1)},{\cal M}_{(2)}, \ldots\}$
with nonabelian gauge groups $\{\G_{(\gamma)}\}$.  
There are a finite number of (semi-simple) groups $\G$ with dimension below
any fixed bound.  For each fixed $\G$, there are a finite number of
representations whose dimension is below the bound (\ref{eq:bound}) on
the number of hypermultiplets.  Thus,
as  argued in \cite{finite},  any infinite family $\{\G_{(\gamma)}\}$ must include
gauge groups of arbitrarily large dimension.  For any given model in
the family we decompose the semi-simple gauge group into a product of simple
group factors $\G_{(\gamma)}= G_1^{(\gamma)} \times G_2^{(\gamma)} \times \cdots \times G_{k(\gamma)}^{(\gamma)} $.
We divide the possibilities into two cases as in \cite{finite}.

\begin{enumerate}
\item The dimension of the simple factors in the groups $\G_{(\gamma)}$ is
  bounded across all $\gamma$, that is $\dim(G_i^{(\gamma)}) \leq D$ for all $1\leq i \leq k(\gamma)$ for every theory $\M_{(\gamma)}$.  In this case, the number of simple
  factors is unbounded over the family.

\item The dimension of at least one simple factor in $\G_{(\gamma)}$ is
unbounded.  For example, the gauge group is of the form
$\G_{(\gamma)}=SU(N_{(\gamma)})\times \tilde{\G}_{(\gamma)}$, where $N_{(\gamma)}
\rightarrow \infty$.
\end{enumerate}

{\bf Case 1}: In this case we can rule out infinite families for
arbitrary but fixed $T$.  In case 1 there are an unbounded number of simple
factors, but the dimension of each factor $G_i^{(\gamma)}$ is bounded by dim
$G^{(\gamma)}_i \leq D$.  Assume that we have an infinite sequence of models
whose gauge groups have $N_{(\gamma)}$ factors, with $N_{(\gamma)}$
unbounded.  To simplify  notation, we drop the subscript $\gamma$ which indexes theories in the family $\{\M_{(\gamma)}\}$.  We consider one model $\M$ in this infinite sequence, with $N$ factors.  We divide the factors $G_i$ into 3 classes:

\begin{enumerate}
\item {\bf Type Z}: $b_i^2 = 0$
\item {\bf Type N}: $b_i^2 < 0$
\item {\bf Type P}: $b_i^2 > 0$
\end{enumerate}

We begin by recapitulating some simple arguments from
\cite{finite}.  Since the dimension of each factor is bounded,
the contribution to $\nhv$ from $-V$ is bounded below by $-N D$.
For fixed $T$ the total number of hypermultiplets is then bounded
by 
\begin{equation}
H \leq 273-29T + N D \equiv B \sim{\cal O} (N)\,.
% \label{eq:}
\end{equation}
This means that the dimension of any given representation is bounded
by the same value $B$.  The number of gauge group factors $\lambda$
under which any matter field can transform  nontrivially is then bounded by
$2^\lambda \leq B$, so $\lambda \leq{\cal O} (\ln N)$.

Now, consider the different types of factors.  Denote the number of
type N, Z, P factors by $N_{N, Z, P}$, where
\begin{equation}
N = N_N + N_Z + N_P \,.
% \label{eq:}
\end{equation}
We can write the
$b_i$'s in a (not necessarily integral)
basis where $\Omega = {\rm diag} (+ 1, -1, -1, \ldots)$ as
\begin{equation}
b_i = (x_i, \vec{y}_i) \,.
\label{eq:bxy}
\end{equation}
For any type P factor, $|x_i | > | \vec{y}_i |$, so $b_i \cdot b_j >
0$ for any pair of type P factors.  Thus, there are hypermultiplets
charged under both gauge groups for every pair of type P factors.  A
hypermultiplet charged under $\lambda \geq 2$ gauge group factors
appears in $\lambda (\lambda -1)$ (ordered) pairs, and contributes at
least $2^\lambda$ to the total number of hypermultiplets $H$.  Each
ordered pair under which this hypermultiplet is charged then
contributes at least
\begin{equation}
\frac{2^\lambda}{ \lambda (\lambda -1)} \geq 1
% \label{eq:}
\end{equation}
to the total number of hypermultiplets $H$.  It follows that the $N_P
(N_P -1)$
type P pairs, under which at least one hypermultiplet is charged,
contribute at least $N_P (N_P -1)$ to $H$, so
\begin{equation}
N_P (N_P -1) \leq B
% \label{eq:}
\end{equation}
Thus,
\begin{equation}
N_P \leq \sqrt{B} + 1 \sim{\cal O} (\sqrt{N})
 \label{eq:order-np}
\end{equation}
which is much smaller than $N$ for large $N$.  
So most of the $b_i's$ associated with gauge group factors in any
infinite family must be type $Z$ or type $N$.

Now consider type N
factors.  
Any set of $r$ mutually orthogonal type $N$ vectors defines an
$r$-dimensional negative-definite subspace of $\R^{1, T}$.
This means, in particular, that we cannot have $T + 1$ mutually
orthogonal type $N$ vectors.  
If we have $N_N$ type $N$ vectors, we can define a graph whose nodes
are the type $N$ vectors, where an edge connects every two nodes
associated with perpendicular vectors.  Tur\'an's theorem \cite{Turan}
states that the maximum number of edges on any graph with $n$ vertices
which does not contain a subset of $T + 1$ completely connected
vertices is
\begin{equation}
(1-1/T) n^2/2
% \label{eq:}
\end{equation}
where the total number of possible edges is $n (n -1)/2$.  Thus,
applying this theorem to the graph described above on nodes associated
with type $N$ vectors, the number of ordered pairs with charged
hypermultiplets must be at least
\begin{equation}
\frac{N_N^2}{T} -N_N   \,.
% \label{eq:}
\end{equation}
It then follows that
\begin{equation}
N_N \leq \sqrt{TB} + T  \sim{\cal O} (\sqrt{N})\,.
 \label{eq:order-nn}
\end{equation}

Finally, consider type Z factors.  Vectors $b_i, b_j$ of the form
(\ref{eq:bxy}) associated with two type Z factors 
each have $| x_i | = | \vec{y}_i |$ and
have a
positive inner product  unless they are parallel, in which case $b_i
\cdot b_j  = 0$.  Denote by $\mu$ the size of the largest collection of
parallel type Z vectors.  Each type Z vector is perpendicular to fewer
than $\mu$ other type Z vectors, so there are  at least $N_Z (N_Z-\mu)$
pairs of type Z factors under which there are charged hypers.  We must then
have
\begin{equation}
N_Z (N_Z-\mu) = (N_Z-\mu) (N -N_P-N_N) \leq  B \,.
% \label{eq:}
\end{equation}
But from (\ref{eq:order-np}, \ref{eq:order-nn}) this means that
$N_Z-\mu$ is of order at most ${\cal O} (1)$ (and is bounded by  $D$ as
$N \rightarrow \infty$), while $N_Z$ is of
order ${\cal O} (N)$.  Thus, all but a fraction of order $1/N$
of the type Z factors have vectors in a common parallel direction.  In
\cite{finite}, we showed that all group + matter configurations
which give type Z factors have a positive value for $\nhv$.  The total
contribution to $\nhv$ is then bounded by
\begin{equation}
\nhv  > \mu - \left[(N_Z-\mu) + N_P + N_N \right] (D)
\sim {\cal O} (N)
% \label{eq:}
\end{equation}
which exceeds the bound $H -V \leq 273-29T$ for sufficiently large
$N$.  Thus, we have ruled out case 1 by contradiction for all $T > 0$.

\vspace*{0.1in}

{\bf Case 2}:

In \cite{finite} we proved that there are no infinite families with
factors of unbounded size for $T = 1$.  A very similar proof works up
to $T = 8$; we outline this proof using the inner product structure
and vectors $a, b_i, j$, making use of results from \cite{finite}.
As discussed in \cite{finite},  for $SU(N)$ the $F^4$ anomaly
cancellation condition
\begin{eqnarray}
B_{\rm Adj} =2N = \sum_R x_R B_R \label{eq:SU-trace}
\end{eqnarray}
can only be satisfied at large $N$ when the number of multiplets $x_R$
vanishes in all representations other than the fundamental, adjoint,
and two-index antisymmetric and symmetric representations.  For these
representations, indexed in that order, (\ref{eq:SU-trace}) becomes
\begin{eqnarray}
2N = x_1+2N x_2+(N-8)x_3+(N+8)x_4 \,.  \label{eq:SU-trace-1}
\end{eqnarray} 
Note that we do not distinguish here between representations and their conjugates, which give equal anomaly contributions.
The solutions to (\ref{eq:SU-trace-1})  at large $N$, along with the
corresponding solutions for the other classical groups $SO(N), Sp(N)$
are listed in Table~\ref{table:solutions}.  We discard solutions
$(x_1, x_2, x_3, x_4) = (0, 1, 0, 0)$ and $(0, 0, 1, 1)$, where $a
\cdot b_i = b_i^2 = 0$ since for $T < 9$ these relations combined with
$a^2 > 0$ imply that
$b_i = 0$ and therefore that the kinetic term for the gauge field is
identically zero.
\begin{table}
\centering
	\begin{tabular}{|c|c|c|c|c|}
	\hline
	Group & Matter content & $\nhv$ & $a \cdot b$ & $b^2$ \\
	\hline
	\multirow{4}{*}{$SU(N)$} & $2N\ {\tiny\yng(1)}$ & $N^2+1$& 0
	& -2\\
	& $(N+8)\ {\tiny\yng(1)}+1\ {\tiny\yng(1,1)}$ & $
	\frac{1}{2}N^2+\frac{15}{2}N+1 $& 1 & -1\\
	& $(N-8)\ {\tiny\yng(1)}+1\ {\tiny\yng(2)}$ & $
	\frac{1}{2}N^2-\frac{15}{2}N+1 $& -1 & -1\\
	& $16\ {\tiny\yng(1)}+2\ {\tiny\yng(1,1)}$ & $ 15N+1 $ & 2 & 0\\
	\hline
	$SO(N)$ & $(N-8)\ {\tiny\yng(1)}$ &
	$\frac{1}{2}N^2-\frac{7}{2}N$ & -1 & -1 \\
\hline
\multirow{2}{*}{$Sp(N/2)$} & $(N+8)\ {\tiny\yng(1)}$ &
	$\frac{1}{2}N^2+\frac{7}{2}N$ & 1 & -1 \\
	 & $16\ {\tiny\yng(1)}+1\ {\tiny\yng(1,1)}$ & $15N-1$ & 2 & 0
	 \\ \hline \end{tabular} \caption{Allowed charged matter for
	 an infinite family of models with gauge group $H(N)$.  The
	 last two columns give the values of $ a \cdot b, b^2$ in the
	 factorized anomaly polynomial.}  \label{table:solutions}
\end{table}
The contribution to $\nhv$ from each
of the group and matter combinations in Table~\ref{table:solutions}
diverges as $N \rightarrow \infty$.  This cannot be cancelled by
contributions to $-V$ from an infinite number of factors, for the
same reasons which rule out case 1.  Thus, any infinite family must
have an infinite sub-family, with gauge group of the form
$\hat{H}(M)\times H(N) \times \tilde{G}_{M, N}$, with both $M,
N\rightarrow \infty$.  For any factors $G_i, G_j$ with $a \cdot b_i, a
\cdot b_j \neq 0$, in a (non-integral) basis where $\Omega = {\rm
diag} (+ 1, -1, -1, \ldots)$, and $a = (\sqrt{a^2}, 0, 0, \ldots)$
writing
\begin{equation}
b_i = (x_i, \vec{y}_i) 
\label{eq:bxy-2}
\end{equation}
with $x_i = a \cdot b_i/\sqrt{a^2}$ we have 
\begin{equation}
x_ix_j = (a \cdot b_i) (a \cdot b_j)/a^2\geq b_i \cdot b_j  = \sum_{R, S}x_{R S} A_R A_S \,.
% \label{eq:}
\end{equation}
Since $x_ix_j$ can be taken to be constant for the infinite family of pairs
$\hat{H} (M), H (N)$, while
$A_R$ grows for all representations besides the fundamental, the only
possible fields charged under more than one of the infinite factors in
Table~\ref{table:solutions} are bifundamentals.

We now consider all possible infinite families built from products of
groups and representations in Table~\ref{table:solutions} with bounded $\nhv$.  There are 5 such
combinations with two factors.  These combinations were enumerated in
\cite{finite}, and are listed in Table 4 in that paper.  These
combinations include two infinite families shown to satisfy anomaly
factorization by Schwarz \cite{Schwarz}, as well as three other
similar families.  In \cite{finite} it was shown that for $T = 1$ the
models in all five of these infinite families are unacceptable because
the gauge kinetic term for the two factors are opposite in sign and
therefore one is always unphysical.  The same consequence follows as long as $T < 9$, where $a^2 > 0$.  This can be shown as follows: For each
two-factor infinite family we have two vectors $b_1, b_2$ which
satisfy $a \cdot (b_1 + b_2) = 0$ and $(b_1 + b_2)^2 = 0$.  But these
conditions imply $b_1 + b_2 = 0$, so that $j \cdot b_1$ and $j \cdot
b_2$ cannot both be positive.  For example, for the theory found by
Schwarz with gauge group $SU(N) \times SU(N)$ with two bifundamental
fields, we have $a\cdot b_1 = a \cdot b_2 = 0, b_1^2 = b_2^2 = -2, b_1
\cdot b_2 = 2$, from which it follows that $b_1 = -b_2$.  This proof
breaks down when $a^2 \leq 0$, since then $a \cdot b = b^2 = 0$ is not
sufficient to prove $b = 0$.  In the following section we give an
infinite family of examples which has no clear inconsistency at $T =
9$.

Note that while for $T = 1$ there are no infinite families with more
than two large gauge factors, at larger $T$ there are families with
three large gauge factors.  For example, there is an infinite family
of models with
\begin{equation}
G = SU(N -8) \times SU(N) \times SU(N+ 8)
% \label{eq:}
\end{equation}
with bifundamental matter in the $(N -8, N, 1)$ and $(1, N, N + 8)$ representations.
(This model cannot occur at $T = 1$ since then it is not possible to
have $b_1 \cdot b_3 = 0$ when $a \cdot b_i = 0, b_i^2 = -2$).  For
this model, and for the similar models with the first and/or last
factor replaced with $Sp(N/2 -4)$ and/or $SO(N + 8)$, a similar
argument to that used to rule out the two-factor infinite families
shows that $b_1 + b_2 + b_3 = 0$ when $T < 9$ so that we cannot have
$j \cdot b_i > 0$ for all three gauge group factors.

This proves case 2 of the analysis.  So we have proven that for $T< 9$
there are a finite number of distinct gauge groups and matter content
which satisfy anomaly cancellation with physical kinetic terms for all
gauge field factors.  We have ruled out infinite families with
unbounded numbers of gauge group factors at any finite $T$, though we
show that such infinite families exist when $T$ is unbounded in
Section \ref{sec:examples-infinite}.  We have not ruled out infinite families
with a finite number of gauge group factors which become unbounded at
finite $T > 8$.  Indeed, we give an explicit construction of such a
family in Section \ref{sec:examples-infinite}.
\vspace*{0.1in}

A systematic enumeration of the finite set of possible gauge groups
and matter content compatible with a fixed $T < 9$ is in principle
possible.  One approach to the enumeration is to break the gauge group
into blocks associated with simple
factors and their associated matter
content, and then to combine blocks in such a way that the
gravitational anomaly $\nhv$ is not exceeded.  This approach was
discussed and applied for some classes of models in \cite{KMT}.  With
more tensor fields, the limit $H-V = 273-29T$ more strictly constrains
the range of possible matter representations, although the increased
dimensionality of the space $\R^{1, T}$ allows blocks to be combined
more freely.  One could proceed with a systematic enumeration by
sequentially classifying all models with matter transforming under at
most $\lambda$ distinct gauge group factors for increasing values of
$\lambda$.  It is easy to see that for any given gauge group there are
a finite number of matter representations such that $H/\lambda -V <
273-29 T$.  This bound is a useful guide in constructing all allowed
models, though care must be taken since for any fixed $T$ there can be
a finite number of type N blocks which contribute negatively to
$\nhv$.  We leave a complete and systematic enumeration of the finite
set of possible $T < 9$ models for future work.

\section{Infinite families for $T \geq 9$}
\label{sec:examples-infinite}

In this section we give some examples of infinite families of models
which satisfy anomaly cancellation and admit correct-sign kinetic
terms, when $T \geq 9$.

\subsection{Example: Infinite families at fixed $T \geq 9$}
\label{sec:infinite-9}

From the way in which the finiteness proof breaks down at $T = 9$, it
is fairly straightforward to construct an infinite class of apparently-consistent supergravity models at $T = 9$.  We consider again the
infinite class of models found by Schwarz with gauge group $G = SU(N)
\times SU(N)$ and two bifundamental matter fields, but now with $T >
8$.  For this gauge group and matter
representation,
at $T = 9$ we need vectors $-a, b_1, b_2$ with inner
product matrix
\begin{equation}
\Lambda = \left(\begin{array}{ccc}
 0 & 0 & 0\\0 & -2 & 2\\0 & 2 & -2
\end{array} \right) \,.
% \label{eq:}
\end{equation}
In a basis with $\Omega = {\rm
diag} (+ 1, -1, -1, \ldots)$, this can be realized through the vectors
\begin{eqnarray}
\begin{array}{rcrrrrrrrrrr}
-a & = & (3, & -1, & -1, & -1, & -1, & -1, & -1, & -1, & -1, & -1) \\
b_1 & = & (1, & -1, & -1, & -1, & 0, & 0, & 0, & 0, & 0, & 0)\\
b_2 & = & (2, & 0, & 0, & 0, & -1, &-1, & -1, & -1, & -1, & -1) 
\end{array}
\label{eq:infinite-9}
\end{eqnarray}
This choice of vectors
satisfies the correct gauge kinetic term
sign conditions $j \cdot b_i > 0$ for $j = (1, 0, 0, \ldots)$.  
It is straightforward to construct similar examples for $T > 9$ by
simply adding additional 1's in additional columns for $a$.

We will show in Section \ref{sec:examples} that 
at $T = 9$
these models are
incompatible with F-theory for large enough values of $N$ and thus do
not have any known string realization.

\subsection{Example: Infinite families with unbounded numbers of
  factors}
\label{sec:infinite-e8}

Although we proved that for any fixed $T$ there are no infinite
families with unbounded numbers of factors (case 1), this restriction
does not hold  when $T$ itself is unbounded.  From the gravitational
anomaly condition (\ref{eq:bound}) it would seem that a family with
increasing $T$ is difficult to construct, as the upper bound on $H-V$
becomes increasingly negative.  By choosing gauge factors with minimal
matter, however, we can find anomaly-free models with arbitrarily many
gauge group factors.  This is not possible for most types of gauge
group factors.  For example, for factors of the form $SU(N)$, as noted
in \cite{KMT}, for any number of antisymmetric tensor representations
the $F^4$ condition fixes the number of fundamental representations so
that $H/2-V$ is positive.  Thus, for factors of this form we cannot
build combinations with arbitrarily negative total $H-V$ with matter
transforming under at most two gauge group factors.

To minimize the total $H-V$ we can consider gauge group factors such
as $SO(8)$ and $E_8$, for which no charged matter is needed to satisfy
the anomaly equations.  For a pure $SO(8)$ factor, we have $V = 28$.
Since each such factor is associated with a type N vector with $b_i^2
= -4$, we need an additional tensor in $T$ to accommodate 
a type N vector
perpendicular to the $b$ vectors from all other factors
for each $SO(8)$ factor.  Adding one
$SO(8)$ factor and one tensor to a model contributes a total of
\begin{equation}
\Delta
(H-V + 29T) = 1
% \label{eq:}
\end{equation}
to the total gravitational anomaly, so the number of such factors
which can be added is large but bounded.

The same is not true for $E_8$ factors.  Each such factor has $V =
248$.  Considering only the constraints from anomaly cancellation and
gauge kinetic term sign conditions, we can construct a family of
models with gauge group $G = E_8^k$ and no charged matter.  The
associated vectors $\{-a, \ b_i\}$ satisfy $-a \cdot b_i =-10, b_i^2 = -12, b_i
\cdot b_j = 0, i \neq j$.  For sufficiently large $T$ such vectors can
be found.  For example, when $T = 9 + 8k$, a representation
with the inner product $\Omega = {\rm diag} (+ 1, -1, -1, \ldots)$
is given by
\begin{eqnarray}
\begin{array}{rcrrrrrrrrrrrrrrr}
-a  & = \ (&  3, & -1, & (-1)_4 & , &   (-1)_4& ,& -1,& -1,& -1, & -1, &\cdots ) \\
b_1 & = \ (& -1, & -1, & (-1)_3,  & -3, & (0)_4 & ,& 0,& 0,& 0, & 0, &\cdots ) \\
b_2 & = \ (& -1, & -1, &  0_4 & , &    (-1)_3,  & -3, & 0, & 0, & 0,& 0, & \cdots) \\
\vdots \\
b_k & = \ (& -1, & -1, & 0_4 &, &  \cdots &, & 0_4, & (-1)_3,& -3,& 0,& \cdots ) 
\end{array}
\label{eq:infinite-e8}
\end{eqnarray}
The notation $x_n$ indicates that the entry $x$ repeats $n$ times.  Note that the last $4k+8$ entries of all the vectors $b_1, \ b_2, \cdots , b_k$ are all zero.  This represents an infinite family of models satisfying anomaly
cancellation.  There exists a choice of $j$ such that gauge
kinetic terms for all factors have the correct sign,
\begin{equation}
j = (-|j_0|, 0, 0, \cdots, 0, 1, 1, \cdots, 1), \quad
|j_0| > \sqrt{4k+8} \,,
\end{equation}
where the last $4k+8$ entries in $j$ are 1.  
By choosing $(4k + 8)/3 > | j_0 |,$ we can also arrange for
$-a \cdot j > 0$.
As we show in Section
\ref{sec:examples}, this class of models is nevertheless not compatible with
F-theory and has no known string realization for large enough $k$.

\section{Supergravity models in F-theory}
\label{sec:F-theory}

Based only on the structure of the low-energy supergravity theory, we
have now shown that there are a finite number of possible gauge groups and
matter representations for models with $T < 9$.  Every consistent
supergravity model, furthermore, is
characterized by an integral lattice $\Lambda$.  The next question we
would like to address is which of these models can be realized in
string theory.  By determining the subset of apparently-consistent
low-energy theories which can be realized through each of the known
approaches to string compactification, we can hope to chart the full
space of 6D supergravity theories.  Identifying characteristic
features of models which cannot be realized through any existing string
construction may lead to the identification of new string vacua, or
new constraints on the low-energy theories.

We focus here on identifying F-theory constructions of low-energy
supergravity models.  In subsection \ref{sec:mapping} we show that the
structure of anomaly-free ${\cal N} = 1$ 6D supergravity theories is
closely related to that of F-theory compactifications, allowing us to
map the discrete data of the 6D supergravity theory to topological
data for an F-theory construction.    This generalizes the analysis of \cite{KMT}, in
which the map from low-energy supergravity to F-theory topological
data was described for models with $T = 1$.

In subsection \ref{sec:geometry}
we examine some of the constraints
on low-energy theories which must be satisfied for an F-theory
realization to exist.
While, as discussed in \cite{KMT}, a large fraction of the apparently-consistent supergravity models at $T = 1$ seem consistent with
F-theory, the constraints imposed by F-theory limit the
range of possible models substantially as $T$ increases.  For $T > 8$,
F-theory reduces the infinite number of apparently-consistent models
to a finite number.    We do not attempt to give a complete and
definitive analysis of the constraints from F-theory on low-energy
theories here, but we identify a number of general constraints on the
structure of the low-energy theory imposed by F-theory.
We give some specific examples
of these F-theory constraints on apparently-consistent models with various
values of $T$ in
Section \ref{sec:examples}.

\subsection{Mapping to F-theory}
\label{sec:mapping}

F-theory\footnote{See \cite{Denef-F-theory} for a good general review
and introduction to F-theory.} \cite{Vafa-f, Morrison-Vafa} is a limit of string
theory which generalizes type IIB string theory by allowing the
axiodilaton $\tau$ to vary over a $d$-dimensional compactification
space $B$.  This can be thought of as describing an auxiliary 2-torus
whose complex structure depends upon the axiodilaton, giving an
elliptic fibration over $B$.  The elliptic fiber degenerates at
complex codimension one loci in the base $B$, which correspond to
7-branes.  Specific types of singularities of the elliptic fibration
structure on divisors $\xi_i$ in the base give rise to corresponding
nonabelian gauge group factors $G_i$ in the resulting low-energy
gravity theory.  When $B$ is a (complex) 2-dimensional space, the
F-theory construction is characterized by an elliptically fibered
Calabi-Yau 3-fold with section. As
shown in \cite{Vafa-f, Morrison-Vafa}, F-theory can be used to describe
nonperturbative string vacua which are inaccessible by direct
supergravity compactification, including 6D models with multiple
tensor fields.  
F-theory is the most general approach developed so far to  construct compactifications of string
theory to six dimensions. There are heterotic
compactifications on K3 with certain kinds of bifundamental matter fields, however, 
which are not described in the standard F-theory approach. We
discuss this further in Section \ref{sec:global};
we are not aware of any other string constructions
of ${\cal N} = 1$ 6D supergravity models which do not also have F-theory descriptions.

In low-energy theories that arise from F-theory compactifications,
various aspects of the low-energy physics including the gauge group
and matter content are controlled by the geometry of the elliptic
3-fold.  
Much of the work on F-theory has focused on understanding the
consequences for the low-energy theory of specific geometric
structures in the Calabi-Yau compactification space.
We would like to turn this around and ask --- Given the
low-energy theory, what are the necessary conditions for the existence
of a UV-completion in the form of an F-theory compactification?

The structure of the integral lattice $\Lambda$ determined by the
vectors $a, b_i$ is closely related to the cohomology lattice of a
two-dimensional F-theory base $B$.  For example, the anomaly conditions
(\ref{eq:bij-condition}) relate the inner product $b_i\cdot b_j$ 
to the number of hypermultiplets simultaneously charged under
$G_i\times G_j$.  In F-theory, the number of such hypermultiplets is
related to the intersection product of divisors $\xi_i$ in the
base $B$ associated
with the nonabelian gauge group factors $G_i$.  Analysis of anomaly
cancellation in F-theory constructions shows that these inner products
can be identified \cite{Sadov, Grassi-Morrison, Grassi-Morrison-2}
\begin{equation}
b_i \cdot b_j = \xi_i \cdot \xi_j \,.
% \label{eq:}
\end{equation}
In fact, as discussed in \cite{KMT},
we can associate the $SO(1,T)$ vector $b_i$ for each factor $G_i$
with a corresponding divisor $\xi_i$ in the base $B$.  This
furthermore leads us to interpret the bilinear form
$\Omega_{\alpha\beta}$ as the intersection product in $H^2(B,\Z)$.
The vector $a$ with norm $a\cdot a = 9-T$ is naturally identified with
the canonical divisor of the base $K_B$, which also satisfies $K_B\cdot
K_B=9-T$.  The inner products between $a$ and $b_i$ also agree with the
corresponding inner products in F-theory, $-a \cdot b_i = -K_B \cdot
\xi_i$ \cite{Sadov, Grassi-Morrison, Grassi-Morrison-2}.  
The requirement of physical
gauge kinetic terms requires that there exist an $SO(1,T)$ vector $j$
satisfying $j\cdot j =1$ with $j\cdot b_i > 0$ for all $i$.  This
vector corresponds in F-theory to the K\"ahler form $J$ on the base
$B$, and the condition $J \cdot  \xi_i > 0$ is the requirement that the
curves wrapped by 7-branes have positive volume.  This successfully
identifies all the parameters of the low-energy theory, up to
the two-derivative level\footnote{except for the metric on the hypermultiplet moduli space}, with geometric quantities in the F-theory
compactification.  To summarize, for any supergravity model with an
F-theory realization, we must have a lattice embedding
\begin{eqnarray}
\Lambda & \hookrightarrow & H^2 (B,\Z) 
\label{eq:map}
\end{eqnarray}
which can be associated with an explicit map from the vectors $a, b_i,
j$ into divisor classes in $B$ so that
\begin{eqnarray}
a & \rightarrow &  K_B \label{eq:a-map}\\
b_i & \rightarrow &  \xi_i\label{eq:b-map}\\
j & \rightarrow &   J\label{eq:j-map}
\end{eqnarray}
where $K_B$ is the canonical divisor, the $\xi_i$ are {\it effective}, 
{\it irreducible} curves, and $J$ is a K\"ahler class on the base.
\vspace*{0.1in}

{\it Example:} $T = 1$

In \cite{KMT} we gave an explicit formulation of the map
(\ref{eq:map}) for the case $T = 1$. 
In that case, the F-theory base manifold is restricted to be a 
Hirzebruch surface
$\F_m$, whose second cohomology admits a basis (integral for even $m$)
\begin{eqnarray}
e_1 & = &  D_v + \frac{m}{2} D_s\\
e_2 & = &  D_s
\end{eqnarray}
in which the intersection form takes the form
(\ref{eq:canonical-omega}).  In terms of the coordinates $\alpha_i,
\tilde{\alpha_i}$ for $b_i$, the divisor associated with each vector
$b_i$ then becomes
\begin{equation}
b_i \rightarrow\frac{\lambda_i}{2} \left( \alpha_i e_1 + \tilde{\alpha_i} e_2
\right) \,.
% \label{eq:}
\end{equation}

The correspondence described through this map
gives us an explicit construction of the topological F-theory data
for any supergravity model which can be realized in F-theory.  Not all
gauge groups and matter representations  associated with
anomaly-free supergravity models, however, have valid F-theory 
realizations.  A complete description of the necessary and sufficient 
conditions on the low-energy
supergravity data which guarantee the existence of an F-theory
construction is somewhat complicated and is left for future work.  We
now describe, however, some of the simple constraints necessary for a
model to have an F-theory realization.  As we discuss in Section
\ref{sec:examples}, these constraints are sufficient to rule out a
number of apparently-consistent models at $T < 9$ and all infinite
families of apparently-consistent models at $T \geq 9$.

\subsection{F-theory constraints on low-energy supergravity}
\label{sec:geometry}

\vspace*{0.1in}

\noindent {\bf Lattice embedding}
\vspace*{0.1in}

The first condition which is clearly necessary to realize a model in
F-theory is the embedding condition (\ref{eq:map}), which states that
the lattice $\Lambda$ must admit an embedding into $H^2 (B, \Z)$ for
some F-theory base $B$. The space $H^2(B, \Z)$, as we discuss in
detail below, has the structure of a unimodular lattice. The embedding
condition (\ref{eq:map}) thus implies the existence of a lattice
embedding of $\Lambda$ into a unimodular lattice.

More specific constraints on which models can be mapped to F-theory
can be determined by giving a complete categorization of the
cohomology groups of complex surfaces $B$ which are acceptable base
manifolds for an F-theory compactification.  To this end
we now discuss in slightly more detail
the geometry of the base $B$, which is a general complex, K\"ahler,
2-dimensional surface with an effective anti-pluricanonical divisor.
(The existence of such a divisor is a weak form of ``positive curvature''.)
The space
$H^2(B,\Z)$ has the structure of a free $\Z$-module (without
torsion) of rank $b_2(B)$, and the intersection product
defines a symmetric inner product, making this into a ``lattice''. Poincar\'e duality further implies that the lattice with the inner product is self-dual, or equivalently unimodular. 
The signature of the lattice is
$(2h^{2,0}+1,h^{1,1}-1)$ by the Hodge index
theorem, where $h^{i,j}$ denote the Hodge numbers.  
If the base $B$ had any holomorphic $1$-forms or holomorphic $2$-forms,
then the total space of the elliptic fibration would also have 
holomorphic $1$-forms of holomorphic $2$-forms, and so it would have
enhanced supersymmetry 
(and necessarily be of the form $(K3\times T^2)/G$ or $T^6/G$).
Thus, to ensure that the F-theory model has exactly ${\cal N}=1$ 
supersymmetry, we must assume that
$h^{1,0}(B)=h^{2,0}(B)=0$; it follows that the lattice has signature
$(1,h^{1,1}-1)$.

There are two key properties of the base $B$ which lead to a complete
classification: first, the line bundles $\mathcal{O}(-4K_B)$ and
 $\mathcal{O}(-6K_B)$
have sections $f$ and $g$ (which serve as coefficients in the
Weierstrass equation of the F-theory model), 
and second, $h^{1,0}(B)=h^{2,0}(B)=0$.
It then follows from the classification of algebraic surfaces (see for example
\cite{BPV}) that either
$B$ is an Enriques surface, $B=\mathbb{P}^2$,
or $B$ is the blowup of a Hirzebruch surface $\mathbb{F}_m$
in $k\ge0$ points.  A third property---that there is no curve in $B$
along which
$f$ vanishes at least $4$ times and $g$ vanishes 
at least $6$ times (the ``minimal'' property
of a Weierstrass equation)---guarantees that $|m|\le12$
\cite{Morrison-Vafa}.

%Since $c_1(B) > 0$,
%the canonical bundle has no global sections and
%$h^{2,0}=h^{0,2}=0$.  
The number of tensor multiplets in the low-energy
theory is $T=h^{1,1}-1$ \cite{Morrison-Vafa}, and so $H^2(B,\Z)$ is a
$T+1$ dimensional unimodular lattice of signature $(1,T)$.  If the
lattice is even, then by Milnor's theorem \cite{Milnor} we must have $T\equiv
1\mod{8}$ and the lattice is isomorphic (as a $\Z$-module) to $U\oplus
E_8(-1)^{\oplus k}$, where
\begin{equation}
U =\begin{pmatrix}
0 & 1 \\
1 & 0
\end{pmatrix}\,.
% \label{eq:}
\end{equation}
When the lattice is odd, then the lattice is isomorphic to $\Z^{T+1}$
with the inner product $\mbox{diag}\{+1,-1,-1,\cdots, -1\}$.  The only
bases with even lattices are the Hirzebruch surfaces $\F_m$ for even
$m$, with lattice $U$, and the Enriques surface with lattice $U \oplus E_8
(-1)$.  All other possible bases are blowups of Hirzebruch surfaces
$\F_m, |m| \leq 12$, and $\P^2$, all of which lead to odd lattices.

To summarize, given a low-energy theory with $T$ tensor multiplets 
characterized by the lattice $\Lambda$, an F-theory realization 
can only exist
if $\Lambda$ embeds, as a lattice, into a signature $(1,T)$ unimodular 
lattice.

In F-theory, the lattice $H_2(B,\Z) \cong H^2(B,\Z)$ corresponds to the charge 
lattice of BPS states obtained by wrapping D3 branes on curves in $B$. 
This is precisely the charge lattice of BPS 
dyonic string states $\tilde{\Lambda}$, discussed in Section \ref
{sec:lattice}, into which there must be a lattice embedding
$\Lambda \hookrightarrow \tilde{\Lambda}$.  
A similar
unimodularity condition arises in the compactification of the
heterotic string on the torus, from modular invariance of the
world-sheet string theory.  It is interesting to speculate whether
this kind of unimodularity condition may be a general consistency
condition for any quantum 6D supergravity theory.  It may be, for
example, that such a condition is necessary for unitarity of the
theory.  We leave further investigation of this question to future
work.

\vspace*{0.1in}
\noindent {\bf Constraint on canonical class and singular divisors}
\vspace*{0.1in}

F-theory imposes strong constraints on the possible values of $a$.
From (\ref{eq:a-map}), $a$ maps to the canonical class $K_B$ of $B$.
For those surfaces with $H^2 (B,\Z) \cong U$, we can always choose 
a basis so that $a
\rightarrow K_B = (-2, -2)$ as discussed in Section
\ref{sec:anomalies}.   For the Enriques surface, $K_B = a = 0$.
For
all the remaining surfaces, we can choose a basis with respect to
which $H^2 (B,\Z)$ has inner product
$\mbox{diag}\{+1,-1,-1,\cdots, -1\}$, and such that $K_B$ takes the form
\begin{equation}
-K_B = (3, -1, -1, \ldots, -1) \,.
% \label{eq:}
\end{equation}
This imposes substantial constraints on the choice of $a$.  In
particular, $a$ is {\em primitive}\footnote{A lattice vector $v$ is {\em primitive} if $\frac{1}{d} v$ is not a lattice vector $\forall \ d \in \Z, \ |d| \neq 1$.} in all cases with odd lattices.

%The vectors $b_i$ must map to {\em effective irreducible divisors}\/
%$\xi_i$ in any F-theory realization under the map in (\ref{eq:b-map}).
%This too imposes constraints, one of which is a primitivity
%constraint: if $b_i^2<0$ then the vector $b_i$ must be primitive.  To
%see this, let $\xi_i$ be an effective, irreducible divisor with
%$\xi_i^2 < 0$, such that $\xi_i$ is not primitive and therefore
%$\xi_i = n \ \xi'$ for another (integral)
%divisor class $\xi'$ with $n>1$.  $\xi'$ is obviously effective and
%irreducible; if not, $\xi_i$ would not be.  We arrive at a
%contradiction because $\xi' \cdot \xi_i < 0 $ implies that $\xi'$ must
%be a component of $\xi_i$.  As a consequence, the image of $b_i$ under
%the map (\ref{eq:b-map}) must be primitive whenever $b_i^2 < 0$.

The geometry of elliptic fibrations implies that vectors $b_i$ must map to {\em effective irreducible divisors}\/
$\xi_i$ in any F-theory realization under the map in (\ref{eq:b-map}).
This constraint has various consequences, an example of which is the following:

\vspace*{0.05in}
\noindent {\em Claim}: If $b^2 < 0$, then the vector $b$ must be primitive in any F-theory realization.
\vspace*{0.05in}

\noindent
We prove this by contradiction; assume that $b$ maps to an 
irreducible, effective divisor $\xi$, and that $b^2 < 0$ with $b$ a 
non-primitive vector. Then, there exists an integer $n > 1$ such that 
the class $\xi' := \xi/n$ is integral and, therefore, an effective 
divisor. Now, $\xi' \cdot \xi < 0$, and since $\xi$ is an irreducible, 
effective divisor this implies that $\xi'$ must contain $\xi$ as a 
component. This is impossible because it would require the class $ (1/n-
k) \xi$ to be effective, for integers $k \geq 1, \ n \geq 2$. 

These conditions on $a, b_i$ impose further constraints on which
supergravity models can be compatible with F-theory.

\vspace*{0.1in}
\noindent {\bf  Positivity conditions and the K\"ahler and Mori cones}
\vspace*{0.1in}

As noted above, the supergravity constraint that all gauge fields have
kinetic terms of the correct sign, $j \cdot b_i > 0$, has a
corresponding interpretation in F-theory.  In F-theory, the divisors
$\xi_i$ supporting the singularity giving rise to nonabelian gauge
group factors must all be effective and irreducible, from which it
follows that $J \cdot \xi_i > 0$ where $J$ is the K\"ahler form of $B$.  This is an example of an F-theory
constraint with a clear analogue in the low-energy theory.  In
F-theory there is a similar constraint on the (negative of the)
canonical class $-K_B$, so that for all F-theory compactifications
$-K_B \cdot J > 0$.  This constraint has no obvious counterpart in
supergravity.  Note, however, that just as supersymmetry constrains
the action so that the gauge kinetic term is proportional to $-j \cdot
 b_i\, {\rm tr}\, F_i^2$ \cite{Sagnotti}, a similar argument suggests
that the action should have a higher-derivative term proportional to
$j \cdot a \, {\rm tr}\, R^2$.  Such higher-derivative terms can have
sign constraints from causality \cite{Allan-Nima}; we leave a further
exploration of this possible constraint on low-energy models for
further work.

%{\bf leaving following paragraph in for now, may be more information
%than needed}.

To understand the F-theory constraints on $\xi_i$ and $J$ more
clearly, it is helpful to  describe in more detail the structure of
the K\"ahler cone and dual Mori cone.

As discussed in Section \ref{sec:anomalies}, the low-energy theories
have an $SO(1,T)/SO(T)$ moduli space of tensor multiplet scalars and a
hypermultiplet moduli space.  In the F-theory compactifications we are
considering, the $h^{1,1}(B)=T+1$ K\"ahler moduli of the base
correspond to the tensor moduli space (except the overall volume of
$B$, which is in a hypermultiplet), while the complex structure moduli
of the elliptic fibration correspond to the hypermultiplet moduli
space of the low-energy theory.  The K\"ahler metric is completely
specified by a choice of K\"ahler form $J \in H^ {1,1}(B,\R) \cong
H^2(B,\Z)\otimes \R$.  Therefore, $J$ is a vector in the space
$\R^{1,T}$ which can be normalized to satisfy $\rm{Vol}(B)=1=\half
J\cdot J$.  To study the structure of the moduli space, we can imagine
starting with a fixed vector $J$ and looking at the transformations of $\R^{1,T}$
that generate inequivalent K\"ahler forms.  The total space of such
transformations is $SO(1,T)$; $SO(T)$ transformations in the
transverse space orthogonal to $J$ do not change the metric, and so
the moduli space of inequivalent metrics is (locally) parameterized by
$SO(1,T)/SO(T)$.

In general, $B$ has an automorphism group $\widetilde{\Gamma}$ and
some quotient
$\Gamma$ of $\widetilde{\Gamma}$ acts faithfully on $H^2(B,\Z)$.  For example,
if $B=\F_0$ or $B$ is the blowup of $\F_0$ at a single point, then
$\Gamma\cong\Z_2$, while if $B$ is the blowup of $\F_0$ at two distinct
points then 
$\Gamma$ is the dihedral group of order $12$.
Note that $\Gamma$ must be
a subgroup of $\operatorname{Aut}(H^2(B,\Z))$, the automorphism group
of the lattice (which
leaves the inner product invariant).
The subgroup
$\Gamma\subset\operatorname{Aut}(H^2(B,\Z))$ induced by automorphisms
of $B$ introduces a further identification on  the
moduli space, since these just correspond to large
diffeomorphisms.  The moduli space of fixed volume, K\"ahler metrics is then
\begin{equation}
\M_{K} \subset \Gamma \backslash SO(1,T)/SO(T)
\end{equation}
This locally agrees with the structure of the moduli space we see from the
low-energy theory.  The discrete group 
$\Gamma\subset\operatorname{Aut}(H^2(B,\Z))$ corresponds to the
S-duality group; in the $T=1$ cases this was discussed in
\cite{Seiberg-Witten}.

The constraint on $J$ from the F-theory side is that $J$ must lie
within the K\"ahler cone of the base surface.  Since $h^{2,0}(B)=0$,
the {\em Kleiman criterion} \cite{kleiman-ampleness} characterizes the
K\"ahler cone as the set of those $J$ such that (1) $J\cdot J>0$ and,
(2) for all effective divisors $D$, $J \cdot D > 0$.  If we normalize
the volume to $1$, the first condition simply states that $J$ lies in
$SO(1,T)/SO(T)$ as above.  To analyze the second condition, it is
useful to work with the dual of the K\"ahler cone, called the Mori
cone \cite{mori-not-nef}, which is the set of linear combinations
$\sum r_iD_i$ of effective divisors $D_i$ using nonnegative real
coefficients $r_i$.  A K\"ahler class $J$ is outside the K\"ahler cone
if $J \cdot D < 0$ for $D$ an effective divisor.  For such K\"ahler
classes, there is no known F-theory vacuum construction.  Note that
the K\"ahler cone is thus essentially defined in terms of the Mori
cone.

The F-theory constraint that $\xi_i$ is effective and irreducible
implies that $\xi_i$ lies within the Mori cone for $B$.
Thus, it follows from combining this constraint with the definition of
the K\"ahler cone that $J \cdot \xi_i > 0$ for all $i$, though the
Mori cone and K\"ahler cone conditions  taken together are a stronger
set of constraints than this inequality which we can understand from
the low-energy theory.
\vspace*{0.05in}

As an example of the Mori cone and dual K\"ahler cone, consider
the bases $\F_m$, which lead to $T=1$.  For these bases $B$, the set of effective divisors
is generated by $D_v$ and $D_s$, with intersection pairings
\begin{equation}
D_v \cdot D_v = -m, \;\;\;\;\;
D_v \cdot D_s = 1, \;\;\;\;\;
D_s \cdot D_s =  0 \,.
% % \label{eq:}
\end{equation}
Effective divisors corresponding to irreducible curves are given by
\begin{equation}
D_v,\hspace*{0.1in} 
a D_v + bD_s, \; a \geq 0, b \geq ma \,.
\label{eq:f-effective}
\end{equation}
In this case, the Mori cone occupies the first quadrant 
$\{aD_v+bD_s\ |\ a\ge0, b\ge0\}$,
and the dual K\"ahler cone can be described as 
$\{aD_v+bD_s\ |\ a\ge0, b\ge ma\}$.
The volume one classes in the K\"ahler cone are those $aD_v+bD_s$ with $a>0$
for which
\begin{equation}
 b = \frac{ma^2+1}{2a}.
\end{equation}

As another example of a K\"ahler cone, we consider the blowup of $\F_1$ in a single
point away from $D_v$ (which coincides with the blowup of $\P^2$
at two distinct points).  We can take as a basis for $H^{1,1}(B)$
the exceptional divisor $E$ as well as curves $D_v$ and $D_s-E$,
where $D_v$ and $D_s$ are pulled back from $\F_1$.  (Note that $D_s-E$
is the proper transform of that fiber on $\F_1$ which passed through
the point which was blown up.)  In this basis, if we write a putative
K\"ahler class as $J=aE+bD_v+c(D_s-E)$ then
\begin{equation} 
J^2 = -a^2-b^2+2ac+2bc-c^2
\end{equation}
so if we set the volume to one we can solve for $c$:
%\[ 1 = -(a+b-c)^2 + 2ab\]
%and it follows that
%\[ a+b-c = \pm \sqrt{2ab-1}\]
\begin{equation}
 c = a+b\pm\sqrt{2ab-1}.
\end{equation}
We learn that $ab\ge\frac12$; also, since $J\cdot D_s=b$ and $D_s$ is
effective, $a$ and $b$ must be positive.  The set of such volume one
classes can be represented as a double cover of the semi-hyperbola
$ab\ge\frac12$ branched on the boundary.

\begin{figure}
\centering
\includegraphics[scale=0.35]{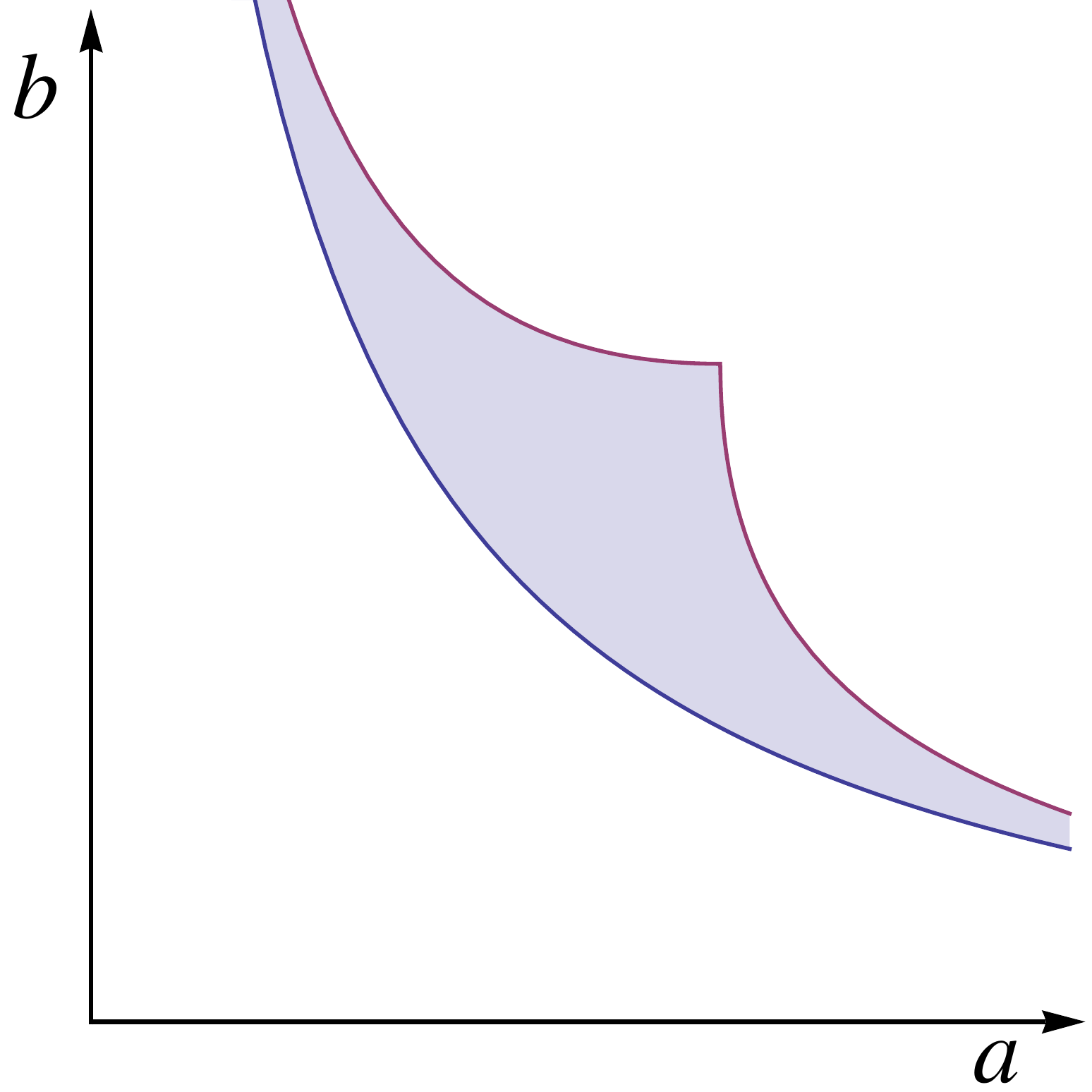}
\caption{K\"ahler cone for the one point blowup of $\F_1$ away from $D_v$.}
\label{fig:Kahler-cone}
\end{figure}

We now investigate the conditions on the K\"ahler cone imposed by
the various effective divisors.  We have 
\begin{align}
J\cdot E &= -a + c = b \pm\sqrt{2ab-1}\\
J\cdot D_v &= -b+c = a \pm\sqrt{2ab-1}\\
J\cdot (D_s-E) &= a+b-c = \mp\sqrt{2ab-1}.
\end{align}
Since all of these must be nonnegative, we must have 
$c=a+b-\sqrt{2ab-1}$
(which selects one of the two branches of the double cover), as 
well as $b\ge \sqrt{2ab-1}$ and $a\ge\sqrt{2ab-1}$.  These latter
two conditions are additional semi-hyperbolas in the $a-b$ plane,
and the region they define is illustrated in Figure \ref{fig:Kahler-cone} 
above.

Note that the $\Z_2$ automorphism in this example exchanges the curves
$D_v$ and $E$, and so acts to exchange $a$ and $b$ in the equations and figure
above.

\vspace*{0.1in}
\noindent {\bf Kodaira condition}
\vspace*{0.1in}

The Kodaira condition corresponds to the mathematical requirement that
the elliptic fibration over $B$ with singularities on the divisors
$\xi_i$ associated with nonabelian group factors gives a Calabi-Yau
manifold.  This condition states that
\begin{equation}
-12K_B = \sum_{i}\nu_i \xi_i + Y
% \label{eq:}
\end{equation}
where $K_B$ is the canonical divisor,  $\nu_i$ are multiplicities
associated with different singularity types ({\it e.g.} $N$ for
$SU(N)$, 6 for $SO(8)$, 10 for $E_8$, etc.), and $Y$ is a residual
divisor, which must be a sum of effective divisors and satisfies $J
\cdot Y > 0$ unless $Y = 0$.  Pulling this relation back to the low-energy theory,
this condition becomes
\begin{equation}
j \cdot (-12a-\sum_{i}  \nu_i b_i)  \geq 0.
\label{eq:Kodaira}
\end{equation}
This condition is also helpful as a simple guide in determining which
low-energy theories admit an F-theory realization.

\vspace*{0.1in}

\noindent {\bf Weierstrass model}
\vspace*{0.1in}

As far as we know, the existence of a map (\ref{eq:map}) to a
prospective F-theory base does not guarantee that there is an F-theory
construction for a given model, even when all the topological and
K\"ahler cone conditions just listed are
satisfied.  An explicit elliptic fibration must be constructed with
correct singularity types for the desired matter content; this is
generally accomplished using Weierstrass models, the Kodaira
singularity classification, and the Tate algorithm
\cite{Morrison-Vafa, Bershadsky-all}.  We do not know any general way to
prove from the topological data that there exists a Weierstrass model
with the desired properties.  In \cite{KMT} we gave explicit
constructions of Weierstrass models for some simple low-energy gauge
group and matter combinations with $T = 1$.  We found that in these
cases there is a precise correspondence between the number of degrees
of freedom needed in the Weierstrass model to construct given gauge
and matter content and the associated value of $\nhv$ in the
low-energy model.  Based on this correspondence we conjectured (by
a simple degree of freedom counting) that it will always be possible to
construct Weierstrass models precisely when the gravitational anomaly
bound on $\nhv$ is not exceeded.  It would be nice to have either a
proof of this conjecture, or an explicit counterexample.  Note,
however, that there are various matter combinations found in
\cite{KMT} which do not currently have known F-theory realizations through
explicit local Weierstrass constructions.  A proof of the general
conjecture that Weierstrass models exist for any model admitting a map
(\ref{eq:map}) satisfying the various topological F-theory conditions would presumably require a
more complete understanding of the range of possible matter content
which can be produced by local singularities, even in the case $T = 1$.

\section{Examples}
\label{sec:examples}

In the previous section we have described some features of the
geometrical data underlying any F-theory construction.  We do not have
a completely determined set of rules which can be used to identify the
subset of low-energy supergravity theories which do have F-theory
realizations.  What does seem to be the case, however, is that when no
map of the form (\ref{eq:map}) exists from $\Lambda, a, b_i, j$ into
the geometry associated with any possible F-theory base $B$ for a
given low-energy model, or one of the previously listed constraints
such as the Kodaira condition is violated, there is no known string
construction of that low-energy model.  Models which do not admit such
a map must at the present time be regarded as lying in the
``swampland'' \cite{swamp} consisting of theories which cannot be
ruled out from low-energy considerations and yet which cannot be
realized in string theory.  In this section we describe some explicit
examples of such low-energy models with no known string theory
realization.

We focus in this section on explicit examples of low-energy theories
which illustrate various features of the associated lattice and
F-theory map.  We begin with several examples encountered in the
previous paper \cite{KMT} on $T = 1$ models.
We first describe a large class of simple models which
do apparently admit F-theory realizations, and then describe
models where the F-theory map 
does not lead to acceptable divisors.  The lattice embedding language
clarifies the issues involved in these cases.  We then return to the
infinite families described in Section \ref{sec:examples-infinite}, and show
how these families violate some of the
conditions for F-theory constructions.

\subsection{Examples with F-theory realizations: $SU(N)$ product models at $T = 1$}

In \cite{KMT} we systematically analyzed a simple general
class of supergravity models, considering all models with gauge group
factors $SU(N), N \geq 4$ and matter in fundamental, bifundamental,
and antisymmetric tensor representations.  We identified all 16,418
models of this type which satisfy anomaly cancellation at $T = 1$.  We performed some basic checks which indicated
that the topological conditions such as the Kodaira condition are
satisfied automatically for all these models as a consequence of the
anomaly cancellation conditions.  We developed Weierstrass models for
a few of these theories and found that generally the number of degrees
of freedom fixed in the Weierstrass polynomial description matched
precisely with the contribution to $H-V$ from each part of the model.
This is expected, as the number of unfixed degrees of freedom
corresponding to moduli in the model should correspond to the number
of uncharged hypermultiplets.  This agreement makes it seem plausible
that all the models in this class, which satisfy all the F-theory
topological conditions, have true F-theory realizations through
Weierstrass models.  As we discuss below, the close agreement between
supergravity anomaly constraints and topological F-theory constraints
seems particularly strong at $T = 1$; for larger values of $T$, we
find more apparently satisfactory low-energy supergravity models which
cannot be realized in F-theory; examples of this type even arise among the class of models with
$SU(N)$ gauge group factors restricted to matter in fundamental,
bifundamental, and antisymmetric tensor representations.

\subsection{Example: Embedding failure}

In \cite{KMT} we found a number of models at $T = 1$ which do not
seem to have acceptable F-theory counterparts.  It is illuminating to
consider these models from the point of view of the lattice embedding
$\Lambda \rightarrow H^2 (B,\Z) $.  One problematic model encountered
in \cite{KMT} has the following structure:

\begin{eqnarray}
G=SU(4) &  &  {\rm matter}= 1 \times {\rm adjoint} + 
10 \times \tiny\yng(1,1)
+40 \times \tiny\yng(1)\\
\Lambda & = &  \left(\begin{array}{cc}
8 & 10\\
10 & 10
\end{array} \right)\\
H-V  &= &220
\end{eqnarray}

It is not hard to check that this lattice $\Lambda$ cannot be
integrally embedded in any unimodular $SO(1, 1)$ lattice.  In
particular, there is no embedding in the lattice $U$, where we can
choose $-a = (2, 2)$; then in terms of (\ref{eq:ba}) we have
$\alpha, \tilde{\alpha} = 5 \pm \sqrt{5}$.  There is also no integral
embedding in the lattice diag $(+1, -1)$.  Choosing $-a = ( 3, -1)$, we
have $b = (x, y)$ with $3x+y = 10, x^2 -y^2 = 10$, with no integer
solutions.

Thus, this model seems not to have a realization in F-theory.  A
similar situation arises for the same group with one adjoint, 11
antisymmetric and 44 fundamental representations, with $\nhv = 242$.
These models are the only ones we have encountered explicitly at $T =
1$ which do not have a unimodular lattice embedding.  It would be very
interesting to understand whether there is a novel string construction
which could lead to models such as these, or if the absence of a unimodular
embedding can be related to a breakdown of consistency for a general
quantum gravity theory.

\subsection{Example: Outside the Mori cone}

Another class of models we found in \cite{KMT} which does not appear
to admit an F-theory realization has less extreme problems. For example, consider a model with

\begin{eqnarray}
G=SU(N) &  &  {\rm matter}= 
1 \times \tiny\yng(2)
+(N -8) \times \tiny\yng(1)\\
\Lambda & = &  \left(\begin{array}{cc}
8 & -1\\
-1 &  -1
\end{array} \right)
\end{eqnarray}

This model has an embedding into the lattice diag $(+1, -1)$ realized
through the vectors $-a = (3, -1), b = (0, -1)$.  If we identify the
lattice diag $(+1,-1)$ with the second cohomology of $\F_1$ by
using the standard basis $(D_u, D_v)$ with $D_u=D_v+D_s$
%for that cohomology
(which satisfies $D_u^2 = 1, D_v^2
= -1, D_v \cdot D_u = 0$),
then the class $-a$ maps to $3D_u-D_v=2D_v+3D_s$ which coincides
with $-K_{\F_1}$.  However, $b$ maps to $-D_v$, which is not an
effective class, so this embedding does not send the vector $b$
into the Mori cone.  
In fact, further checking shows that {\em no}\/
embedding in this case maps $a$ to $K_{\F_1}$ while sending $b$
into the Mori cone.
Thus, there is no F-theory realization of these models.

There are a number of other low-energy models which do not admit an
F-theory realization in the analysis of \cite{KMT}, and which suffer from
similar ``cone'' problems when a lattice embedding is found.  For example,
we encountered in
\cite{K3} a model with structure

\begin{eqnarray}
G=SU(24)  \times SO(8) &  &  {\rm matter}= 
3 \times ( \tiny\yng(1,1),  1)\\
\Lambda & = &  \left(\begin{array}{ccc}
8 & 3 & -1\\
3 & 1 & 0\\
-1 & 0 & -1
\end{array} \right)\\
H-V & = &  225
\end{eqnarray}

This lattice is degenerate and admits an embedding into diag $(+1,-1)$ through
\begin{eqnarray}
-a & = &  (3, -1)\\
b_1 & = &  (1, 0)\\
b_2 & = & (0, -1)
\end{eqnarray}
Just as above, this is compatible with the intersection form on
$\F_1$, but where $b_2 \rightarrow -D_v$, giving a  class
outside the Mori cone.

It would be interesting to investigate whether there is some residue
of the Mori cone constraint in low-energy supergravity.  One can 
imagine that a class of apparently-consistent low-energy
theories with no F-theory realization may be given by a novel stringy
construction analogous to F-theory outside the Mori cone (or its
dual, the K\"ahler cone).  It is
familiar from type II compactification that passing outside the
K\"ahler cone may give valid orbifold or non-geometric Landau-Ginzburg
phases of string compactifications \cite{Polchinski2}.  This seems harder to understand
in the F-theory context, where the moduli are real,\footnote{Note that
in another context with real K\"ahler moduli -- compactifications of M theory
to five dimensions -- some of the walls of the K\"ahler cone serve as
obstructions to further deformation \cite{WitMF,delPezzo}.} than in the type
II context, where the moduli are complex and it is easier to
continuously deform a model to a region outside the K\"ahler cone.  
In any case,
it would be interesting to
explore this phenomenon in F-theory, or through dual constructions.

\subsection{Example: New possibilities and constraints for $T > 1$}

When the number of tensor multiplets $T$ increases, the constraint from
the gravitational anomaly becomes stronger as $H-V \leq 273-29T$.
This would seem to more strongly constrain the number of possible
models.  In general, models which are possible at $T = 1$ continue to
be acceptable until $T$ is so large that the number of charged
hypermultiplets exceeds the constraint from the gravitational anomaly;
in most cases this can be realized by simply adding an additional unit
to $-a$ in the extra negative-definite dimension provided by each
additional tensor multiplet.
Because the dimensionality of the space in which $a, b_i$ are
embedded increases as $T$ increases, however, new possibilities for
combinations of gauge group factors arise.  Although this increases
the number of apparently-consistent supergravity models,
F-theory realizations of
these new possibilities at larger values of $T$ are more strongly constrained by the
condition of embeddability in a unimodular lattice.

\vspace*{0.1in}
\noindent {\bf $SU(N)$ factors at $T = 2$}
\vspace*{0.1in}

A simple example of how the situation changes at increased $T$ is
given by the class of models having gauge group factors $SU(N)$ with matter only
in the fundamental representation.  Anomaly cancellation conditions
require that the number of fundamentals for each such factor
is $f = 2 N$.  At $T = 1$,  models with a single $SU(N)$ gauge group factor are possible up to $N = 15$ by
the gravitational anomaly condition.  It seems that these models all
admit F-theory realizations.  The associated divisor in $\F_2$ for
these models is $D_v$, and $-K = 2D_v + 4D_s$, so all topological
constraints are satisfied for each of these models.  We constructed
explicit Weierstrass models for this family of models up to $SU(14)$ in \cite{KMT},
and believe that such a model exists for $SU(15)$, though the algebra
needed to explicitly construct such a solution
becomes complicated.

Now consider models with multiple $SU(N)$ factors under each of which all matter transforms under either the trivial or fundamental representation.
The vector $b$ associated with an $SU(N)$ factor
with $f = 2 N$ fundamental matter fields must satisfy $a \cdot b =0,
b^2 = -2$.  
At $T = 1$, we cannot have a model with more than one such $SU(N)$
factor and only fundamental matter, since $a \cdot b = 0$ gives a
unique vector $b$ up a scale factor.
When we consider analogous models at $T = 2$, however, the situation
is rather different.
Consider a group $G = SU(N) \times SU(N)$ where each of the
matter fields transforms in the fundamental representation of at most one
$SU(N)$ factor (we assume for simplicity that there are no bifundamental matter
fields).  For $T = 1$, as just mentioned, there is no consistent model with
physical gauge kinetic terms.
On the other hand, when $T > 1$, we have the same inner products $a
\cdot b_i, b_i \cdot b_j$, but the vectors $b_1, b_2$ need not be
linearly dependent.  The associated lattice is 
\begin{equation}
\Lambda  =   \left(\begin{array}{ccc}
9-T & 0 & 0\\
0 & -2 & 0\\
0 & 0 & -2
\end{array} \right)
% \label{eq:}
\end{equation}
From the low-energy point of view this seems like a perfectly
acceptable model.  There is, however, no embedding of this lattice
into a unimodular lattice for $T = 2$.  
In the canonical form of the $SO(1, 2)$ metric
diag $(+1,-1, -1)$, where $ -a = (3, -1, -1)$, if we write $b = (x, y, z)$
for integral $x, y, z$
then we must have
\begin{eqnarray}
3x + y + z & = &0 \\
x^2 -y^2 -z^2 & = &  -2\,.
\end{eqnarray}
The only solutions to these Diophantine equations are
\begin{equation}
b = (0, 1, -1), \;\;\;\;\; b = (0, -1, 1) \,,
% \label{eq:}
\end{equation}
so we cannot have two  vectors $b_1, b_2$ in a (1, 2) unimodular
lattice which are perpendicular both to one another and to $a$, where
both have norm $-2$.
Thus, though at $T = 2$ this
model seems perfectly acceptable from the low-energy point of view, it
cannot be realized through F-theory by any known mechanism.  

We see from this example that as $T$ increases, the constraints on
which gauge factors and matter content can be combined in the
low-energy theory are reduced, and the unimodular embedding constraint
becomes a stronger constraint.  This seems to be a general feature of
models at $T > 1$.  Thus, the fraction of models in the
swampland seems to increase at larger $T$, even though the
gravitational anomaly constraint becomes more stringent.

\subsection{Example: Infinite families at  $T  \geq 9$}
\label{sec:infinite-f9}

In Section (\ref{sec:infinite-9}) we described an infinite family of
models at $T \geq 9$ analogous to the $SU(N) \times SU(N)$ models
shown by Schwarz to satisfy anomaly factorization at $T = 1$ in
\cite{Schwarz}.  It is clear from the above discussion that these
models violate the consistency conditions needed for an F-theory
construction.  Note that the $a$ and $b_i$'s from
(\ref{eq:infinite-9}) satisfy $-a = b_1 + b_2$ at $T = 9$.  This means
that $Y =-12a-N (b_1 + b_2)$ must have $j \cdot Y < 0$ at $N > 12$,
violating the Kodaira condition (\ref{eq:Kodaira}).  The vectors
listed in (\ref{eq:infinite-9}) are not uniquely determined, so one
may suspect that there might exist another choice of vectors that satisfy the
necessary conditions for an F-theory embedding.  This is not possible, however.
The preceding argument can be strengthened and generalized along the
following lines to show that no infinite family of models can be
realized in F-theory at $T = 9$.  For any such model we have, as
discussed in case 2 in the proof in Section \ref{sec:finite}, $a \cdot
(b_1 + b_2) = 0$ and $(b_1 + b_2)^2 = 0$.  For $T = 9$, $a^2 = 0$ so
$a$ is a null (type Z) vector.  It follows that $b_1 + b_2 = -x a$.
Moreover, since $-a, b_1, b_2$ are effective, we have $x>0$.  The
vector $a$ must be primitive in F-theory, and can be put in the form $-a = (3, -1, -1, \cdots)$,
so $x$ is an integer.  So, $j \cdot Y = j\cdot (-12a) (1 - a_N x)$,
where $a_N$ is $\frac{N}{12}$ for $SU(N)$, $\frac{N+6}{12}$ for
$SO(2N)$, $\frac{N}{12}$ for $Sp(\frac{N}{2})$ \cite{Morrison-Vafa}.
Since $a_N$ grows with $N$ in all three cases, at large enough $N$ we
must have $j \cdot Y < 0$.  It follows that the Kodaira condition is
violated at large enough $N$.  This bounds the range of $N$
at $T = 9$ for any of
the infinite classes of models considered in case 2 in the proof of
finiteness.  Note that in \cite{Dabholkar-Park}, the $SU(N) \times
SU(N)$ model with 9 tensor multiplets was identified for $N = 8$.

We also described in Section \ref{sec:infinite-e8} an
infinite family of models with gauge group $E_8^k$ and $T = 9+8k$ for
arbitrary $k$.  
For the vectors $-a, \ b_i$ listed in (\ref{eq:infinite-e8}), there is no F-theory realization.  This is because the residual locus in such an embedding is not effective.  It is easily checked that 
\begin{equation}
\begin{aligned}
j\cdot Y & =  12j\cdot (-a) -10 \sum_{i=1}^k j\cdot b_i 
\\&= 12(4k+8) - |j_0| (10k+36) \\
& <  12(4k+8)-(10k+36)\sqrt{4k+8} < 0, \mbox{ for } k \geq 1
\end{aligned}
\end{equation}
This conclusion depends upon the choice of vectors in
\ref{eq:infinite-e8}.
Note that the $k=2$ member of this family can be realized in heterotic-M-theory compactified on $K3\times S^1/\Z_2$, as outlined in \cite{Seiberg-Witten}.
The fact that this infinite family of models cannot be realized in
F-theory also follows,
however, from a more general argument which we now give.

In fact, there is a uniform bound on the rank of the total
gauge group which holds for all F-theory models (although we do not
know precisely
what the bound is), and this excludes the infinite family with gauge group
$E_8^k$ as well as all other infinite families.  To see that there is
such a bound, note that, as discussed previously, the base $B$ of
any F-theory model must admit a map to a minimal surface $B_{\text{min}}$
which is either an Enriques surface, $\mathbb{P}^2$, or
one of the Hirzebruch surfaces $\F_m$
with $|m|\le12$.
The coefficients $f$ and $g$ in the Weierstrass equation of the
F-theory model
push forward to 
sections $\bar{f}\in H^0(-4K_{B_{\text{min}}})$ and 
$\bar{g}\in H^0(-6K_{B_{\text{min}}})$, with
an induced discriminant 
$\bar{\Delta} = \{4\bar{f}^3+27\bar{g}^2=0\}\in |-12K_{B_{\text{min}}}|$.
Now each component of the gauge group is either associated to a component
of $\bar{\Delta}$ (with its rank determined by the multiplicity) or
to a singular point of $\bar{\Delta}$ which is blown up by the map
$B\to B_{\text{min}}$.  The multiplicities of the components of $\bar{\Delta}$
are uniformly bounded (since $-12K_{B_{\text{min}}}$ can be written
as a sum of effective divisors in only finitely many ways), 
so we only need to show that the total ranks of gauge groups
coming from singular points are bounded.  Note that when $B_{\text{min}}$
is an Enriques surface, $\bar{\Delta}$ is empty so there is nothing to 
check.

For each fixed $B_{\text{min}}$ which is not an Enriques surface,
we consider the
set of all possible pairs 
$(\bar{f},\bar{g}) \subset H^0(-4K_{B_{\text{min}}})
\oplus H^0(-6K_{B_{\text{min}}})$.  We can stratify this
set according to the types of singularities of $\bar{\Delta}$ which
appear, and each stratum is a locally closed
algebraic subset.  Moreover, each stratum has a unique associated gauge
group, and so there is a specific rank which is associated to it.

But the Hilbert basis theorem implies that any stratification of an
affine algebraic variety into locally closed algebraic subsets has only
finitely many strata.  Thus, there are only finitely many possible different
gauge groups which can  occur, so in particular, their ranks must be
bounded.  And since there are only finitely many possibilities for
$B_{\text{min}}$ (other than Enriques surfaces), 
there is a uniform bound for all F-theory models.  
As in the proof of finiteness from Section \ref{sec:finite}, this also
shows that there is a finite number of distinct gauge groups and
matter representations which can be realized through F-theory.

\section{Global picture of the 6D ${\cal N} = 1$ landscape}
\label{sec:global}

The strong constraints which anomalies place on nonabelian gauge group
structure and matter content in ${\cal N} = 1$ 6D theories have given
us a global outline of which of these supergravity theories have the
potential to describe consistent quantum theories.  Explicit knowledge of this set of theories gives us a powerful
tool for exploring the connection between string theory and low-energy
physics.  We can in principle make a  list of the finite number of possible
theories with $T <9$.  For each of the various approaches to string
compactification (heterotic, F-theory, \ldots) we can then determine
which subset of these possible theories can be realized through each
class of construction.  For $T > 8$, there are infinite families of
models satisfying the known low-energy consistency conditions, and our
analytic control of the total space is weaker.  In this section we
summarize some of our knowledge regarding the extent and connectivity
of the 6D ${\cal N} = 1$ supergravity landscape.

Note that  the term ``landscape'' is often used to denote a space of
effective theories containing discrete points with no massless moduli
(flat directions).  Generally such a landscape includes supersymmetric
$AdS$ vacua in addition to
metastable $dS$ vacua with broken supersymmetry. 
In the case of
six-dimensional vacua, it is actually impossible to stabilize all the
moduli while preserving supersymmetry.
In Minkowski space, the gravitational anomaly condition 
$H-V+29T=273$ makes it impossible to avoid moduli. Since the tensor 
multiplet contains massless scalars, to avoid tensor moduli we must 
have $T=0$. This implies that $H \geq 273$ and therefore we have 
hypermultiplet moduli.
In six dimensions, there are also no ${\cal N} = 1$ supersymmetric 
$AdS$ vacua. This can be seen from the fact
that there are no $AdS_6$ superalgebras with 8 supercharges in the
Nahm classification \cite{Nahm}.    The ``landscape'' in this paper
refers to the complete moduli space of 6D gravity theories with one
supersymmetry, although all these vacua are Minkowski and have
massless scalar moduli.

\subsection{Extent of the  space of 6D supergravities}

For $T = 1$, our understanding of the space of theories, while still
incomplete in many respects, seems to suggest a simple global picture.
There are a finite number of low-energy nonabelian gauge groups and
matter content which are compatible with anomaly cancellation and
physical constraints on gauge kinetic terms.  We have an explicit
approach to embedding these models into F-theory,
which seems successful for almost all acceptable gauge groups and
matter content.  There are a few exotic combinations of gauge groups
and matter representations which give lattices which cannot be
embedded into any unimodular lattice associated with an acceptable
F-theory base.  For the moment, these models live in the swampland.
In the $T = 1$
supergravity landscape there are also some models which we have found
whose realization in F-theory would require divisors outside the Mori cone.  
This may correspond to a new class of orbifold or
non-geometric F-theory construction whose precise implementation
remains to be elucidated.  For the vast majority of $T = 1$ models, we
have an explicit map from each model to a set of divisor classes in a
given F-theory base.  Degree of freedom counting suggests that this
topological data can be completed to a full F-theory construction
through a Weierstrass model, though a proof of this assertion in a
general context remains to be found.  We
have not addressed here the question of $U(1)$ factors in the gauge
group and their associated charges.  The anomaly constraints on $U(1)$
factors are more complicated \cite{Erler} and lead to systems of
Diophantine equations, whose analysis will be discussed elsewhere
\cite{Park-Taylor}.

We have shown in this paper that the situation for $T < 9$ is 
similar to that for $T = 1$ as $a^2$ is positive.  Again, there are a finite number of
distinct nonabelian groups and matter representations possible for
this class of models.  
As $T$ increases, it seems to be easier to construct models which 
violate the conditions outlined in Section \ref{sec:geometry} and 
which, therefore, have no F-theory
realization.  Thus, the apparent swampland increases as $T$ increases.
%Nonetheless, the structure is qualitatively similar to that of $T = 1$
%at smaller values of $T$, where the range of possibilities is
%controlled by a positive $a^2$.

For $T \geq 9$, however, the situation changes dramatically.  As $a^2$
becomes negative, infinite families of apparently-consistent
low-energy theories appear.  Some infinite families arise at fixed
$T$, such as an infinite family we have explicitly constructed with
gauge group $SU(N) \times SU(N)$ at $T \geq 9$.  The
infinite families at finite $T$ must contain a bounded number of gauge
group factors, as shown in Section \ref{sec:finite}.  Other infinite families of
models extend to arbitrarily large values of $T$, and the number
of simple factors in the gauge group can become unbounded as $T
\rightarrow \infty$.  
The infinite families described in Section \ref{sec:examples-infinite}
satisfy the unimodular embedding constraint, indicating that 
this constraint is  not a strong constraint for the existence of an
F-theory construction.  
As shown in Section \ref{sec:infinite-f9}, however, the infinite
families we have explicitly constructed are not compatible with other
constraints from F-theory, such as the Kodaira constraint.

The bounded rank argument in Section \ref{sec:infinite-f9} shows that the
number of 6D ${\cal N} = 1$ supergravity models compatible with
F-theory must be finite.  A heuristic version of this argument is as
follows: each of the nonabelian gauge group factors $G_i$ and
associated matter fields are realized in F-theory by tuning the
coefficients of polynomials $f, g$ in the Weierstrass model
\begin{equation}
y^2 = x^3 + f x + g \label{eq:Weierstrass} \,.
\end{equation}
For any F-theory model on a blowup of a space $\F_m$, $f, g$ can be
thought of as
sections of $-4K, -6K$ respectively over (a generally singular) $\F_m$, and have a fixed number of total
coefficients available for tuning (roughly 244 $= 273-29$).  Since
each gauge group factor and associated matter require tuning
additional coefficients to achieve the desired singularity type for
the fibration (which
may be at a singular point in the base which needs to be blown up), only a finite
number of distinct combinations of gauge groups and matter content can
be realized in this fashion.

This argument can be translated into the language of the low-energy
supergravity theory.  As described in \cite{KMT}, in any 6D ${\cal N}
= 1$ supergravity theory the set of fields can be decomposed into
``blocks'' associated with the simple factors in the gauge group and
associated matter representations.  The finiteness argument above for
F-theory models suggests that, for models consistent with F-theory,
each block added to a model must contribute positively to $H-V + 29T$.
This is certainly the case for most supergravity blocks associated
with F-theory singularities which we understand.  We leave a more
general and rigorous proof of these assertions for future work, but
this argument suggests that by classifying individual blocks which
contribute positive values $H-V + 29T$ we should in principle be able
to enumerate all of the finite set of supergravity models with
possible F-theory realizations at any fixed $T$, even when exotic
matter fields such as those encountered in \cite{KMT} are included for
which the F-theory singularity type is not yet classified.  Including the
transitions we discuss in the following subsection which change the
value of $T$, corresponding to blowing up singular points on the
F-theory base, would in principle make it possible to connect the
entire finite set of F-theory vacua in terms of the low-energy
structure.

It was recently shown that all 10D supergravity theories not realized
in string theory are inconsistent as quantum theories \cite{adt}.
It was conjectured in \cite{universality} that this ``string
universality'' property holds for 6D ${\cal N} = 1$
supergravities.  Whether  or not this is true,
it is interesting to speculate that the constraints associated with
F-theory constructions may have some shadow in the low-energy theory
which can lead to new quantum consistency conditions for 6D
supergravities.  For example, the constraint that $\Lambda$ be
embeddable in a unimodular lattice, or the sign constraint $-a \cdot
j>0$ may be realizable in some simple way as consistency conditions on
any low-energy supergravity theory, as discussed in Section
\ref{sec:F-theory}.  For other constraints, such as the detailed
constraints on the form of $a$ or the K\"ahler/Mori cone, it is harder to
see how such conditions can arise directly from the supergravity
description.  It may be that these conditions can best be understood
in terms of the BPS string states underlying the anomaly lattice
$\Lambda$; we hope to return to this question in future work.

\subsection{Charting the space of supergravities with string
  constructions}

One important question is whether the constraints imposed by F-theory
are satisfied by all 6D models arising from string constructions, or
whether they are just signatures of an F-theory ``corner'' of the 6D
supergravity landscape.  We have looked at some examples of 6D
theories realized through other string constructions, including CFTs
and Gepner models \cite{Bianchi, Schellekens}, orientifold models
\cite{orientifolds, Dabholkar-Park} intersecting brane models
\cite{Blumenhagen, Nagaoka-Taylor},  heterotic constructions
\cite{gswest, K3}, and non-geometric string vacua \cite{hmw}.  In general, at least from a limited sampling, it
seems that most of the low-energy theories associated with these
constructions can be mapped to acceptable data for an F-theory
construction, so F-theory seems to cover a large fraction of the space
of 6D low-energy theories which can be realized through any string
construction.  It would, however, clearly be desirable to explore the
other branches of string theory more completely, to develop a more
systematic understanding of how the sets of low-energy theories
arising from other string constructions intersect with those coming
from F-theory, and to ascertain whether the constraints described in
\ref{sec:geometry} are truly universal stringy constraints for 6D
${\cal N} = 1$ theories.

One exception we have encountered to the general existence of F-theory constructions is
given by a class of heterotic line bundle constructions described in
\cite{K3}.  In that paper we analyzed a class of low-energy models
arising from heterotic string compactifications which are also
characterized by lattices, but in a slightly different fashion than
those models considered here.  For a fixed gauge group containing
factors $U(N) \times U(M)$, the models examined in \cite{K3} can have
some number of bifundamental matter fields in the $(N, \bar{M}) + (\bar
{N},M)$ representation and some number of bifundamental matter fields in
the $(N, M) +  (\bar{N}, \bar{M})$ representation.  Each type of field
contributes in the same way to the anomaly polynomials, so models with
the same total number of bifundamental fields map through
(\ref{eq:map}) to identical F-theory constructions.  We do not know of
any mechanism in F-theory as it currently exists to construct models
with different distributions of bifundamental matter fields of these
two types.  Thus, in this class of models the heterotic theory
generates models which cannot be realized in F-theory, although these
models do not violate any of the constraints described in Section
\ref{sec:geometry}.  It would be nice to no whether
there is a generalization of F-theory which would capture these
heterotic models with mixed classes of bifundamental fields.

One class of models which we have not yet considered is the class of
gauged supergravity models
\cite{gauged, Suzuki-Tachikawa}.  It may be possible to perform
a similar analysis of general 6D gauged supergravity models, although
the significance of such supergravity theories is unclear as they do not give rise to
stable Minkowski vacua.  We leave this for
future work.

\subsection{Connectivity of the space of 6D supergravities}

In most of this paper, and in the preceding discussion, we have
referred to supergravity models with distinct gauge groups and matter
content as distinct ``theories'' or ``models''.  This is not quite correct.  In
fact, each consistent model with a fixed gauge group and matter
content has a moduli space of vacua.  At certain limits in the moduli
space, the theory can develop a singularity and the field content can
change in a discrete fashion.  The simplest example of this is the
phenomenon of Higgsing, familiar from the standard model and basic
quantum field theory, in which a vacuum expectation value of scalar fields breaks
a gauge symmetry, giving mass to a formerly massless gauge field while
removing one or more scalar fields from the spectrum.  Such
transitions in the full space of six-dimensional supergravity theories
connect different branches while preserving the total $\nhv$.  More
exotic transitions, studied in \cite{dmw,
Seiberg-Witten, Morrison-Vafa}, arise at singular points where strings
become massless, associated with points in the low-energy theory where
$j \cdot b_i$ vanishes.  Passing through such transition points can
change the number of tensor multiplets in the theory, still preserving
$\nhv+29T$.  In this way, the  set of six-dimensional supergravity
models is really a highly-connected space with many
branches of different dimensionality.  The tools we have developed in
this and preceding papers may be useful in exploring the global
structure of this space.  For example, 
the anomaly constraints can be used to characterize  allowed
transitions in terms of the field content of the low-energy theory.
With a better understanding of
what kinds of transitions between branches are allowed, it may be
possible to prove that the space of acceptable models is connected
into a single moduli space.  Thus, we may be able to probe the
validity of at least the elliptically-fibered version of the Reid
fantasy \cite{Reid} by analysis of the connectivity structure of the
countable set of apparently-consistent low-energy ${\cal N} = 1$ 6D
supergravity theories.

If in fact it could be shown that the set of consistent low-energy 6D
theories is connected, it would provide a picture in which there is a
single consistent ${\cal N} = 1$ supergravity theory, with many
connected branches having different gauge groups and matter content.

If it could be shown that either the entire space of theories, or a
connected subset thereof, corresponds to the set of theories which can
be realized by F-theory or other string constructions, it would give a
very simple picture of string theory as a single unified theory
underlying quantum gravity in 6D.  Indeed, it seems likely that this
can be realized for the set of models which can be realized through
F-theory, since the various singularity types realized by tuning the
coefficients of $f, g$ in (\ref{eq:Weierstrass}) are all connected
continuously in the space of coefficients.  Thus, at least the space
of F-theory compactifications should form a connected moduli space,
with various branches associated with 6D supergravity theories
with various gauge groups and matter content.

\section{Conclusions}
\label{sec:conclusions}

In this paper we have addressed some global questions regarding the
space of consistent supergravity theories in six dimensions.  We have
focused on theories with ${\cal N} = 1$ supersymmetry and nonabelian
gauge groups.  We have extended our previous analysis of such theories
to incorporate multiple tensor multiplets.  We have shown that when
the number of tensor multiplets $T$ is less than 9, there are a finite
number of possible gauge groups and matter representations possible
for such theories.  We have identified infinite families of models at
$T = 9$ and greater which satisfy anomaly cancellation and which have
proper signs for all gauge kinetic terms.

We have shown that every consistent 6D supergravity theory can be
associated with an integral lattice, associated with the coefficients
in the anomaly polynomial.  This lattice can be used to construct
topological data for an F-theory compactification whenever one exists.
We have found a variety of low-energy supergravity models which do not
violate any known consistency conditions from the low-energy point of
view but which have no embedding in F-theory.  The geometrical
constraints of F-theory provide criteria for identifying such
low-energy models from the associated lattice structure, and suggest
possible new low-energy consistency conditions for quantum
supergravity theories.

The overall picture is that, while for Lagrangian models with only one
tensor field most apparently-consistent supergravity theories have
realizations in F-theory, for models with more tensor fields the vast
majority of apparently-consistent models have no known string
realization through F-theory or any other string vacuum construction.
Thus, most of these models lie in the ``swampland''.  If these models
cannot be realized through some novel string construction, it will
indicate that string theory imposes strong constraints on 6D ${\cal N}
= 1$ supergravity theories beyond the known stringent anomaly
cancellation and gauge kinetic term sign constraints.  If these
additional constraints can be understood in terms of new quantum
consistency conditions on the set of low-energy effective theories, it
will provide a new window on general theories of quantum gravity; if
not, it will indicate the existence of stringy constraints which may
distinguish string theory from other possible UV-complete quantum
gravity theories.

The perspective and tools developed in this work provide a framework
in which it may be possible to carry out a systematic mapping of the
landscape of 6D supergravity theories.  Identifying which subsets of
this landscape are associated with the different classes of string
vacuum construction, and understanding how the many branches of this landscape
are connected through Higgs and tensionless string transitions, promises
to lead to a richer understanding of how the different string
constructions are related, and of the nature of the landscape and the
swampland.  Such understanding in the simpler case of six dimensions
will hopefully teach us some new lessons which may be relevant in the
more complicated and physically relevant case of four dimensions.
\vspace*{0.05in}

{\bf Acknowledgements}: We would like to thank Allan Adams, Maissam Barkeshli, Massimo
Bianchi, Oliver de Wolfe, Thomas Faulkner, Greg Moore, John McGreevy,
Satoshi Nagaoka, Daniel Park, Sakura Sch\"afer-Nameki, 
John Schwarz and Ergin Sezgin for helpful discussions.  W.T.\ thanks the Kavli
Institute for Theoretical Physics and the Aspen Center for Physics for hospitality during various
stages of this project, and D.R.M.\ thanks the Aspen Center for
Physics and the Simons Center for Geometry and Physics
for hospitality during the final stages.  
V.K.  thanks the Kavli Institute for Theoretical Physics for support 
through the Graduate Fellows program.  This research was
supported by the DOE under contract \#DE-FC02-94ER40818 and by the
National Science Foundation under grants DMS-0606578 and PHY05-51164.  
Any opinions,
findings, and conclusions or recommendations expressed in this
material are those of the authors and do not necessarily reflect the
views of the granting agencies.

\end{document}